\documentclass[reprint,prd,nobibnotes,nofootinbib]{revtex4-1} 
\usepackage{amsfonts}
\usepackage{graphicx}
\usepackage{amsmath}

\begin{document}

\title{Second Order Gravitational Self-Force}

\author{Samuel E. Gralla}
\affiliation{Department of Physics \\ University of Maryland \\
College Park, MD 20742-411 }

\begin{abstract}
The second-order gravitational self-force on a small body is an important problem for gravitational-wave astronomy of extreme mass-ratio inspirals.  We give a first-principles derivation of a prescription for computing the first and second perturbed metric and motion of a small body moving through a vacuum background spacetime.  The procedure involves solving for a ``regular field'' with a specified (sufficiently smooth) ``effective source'', and may be applied in any gauge that produces a sufficiently smooth regular field.
\end{abstract}

\maketitle

With the promise of gravitational-wave astronomy, the self-field corrections to the motion of a small body have left the domain of pure theory to become a topic in astrophysics.  The principle system of interest is a compact object orbiting a supermassive black hole, or ``extreme mass-ratio inspiral''.  To obtain sufficiently accurate gravitational waveforms for data analysis and parameter extraction, one must go beyond the geodesic approximation to include deviations caused by the body's finite mass (e.g., \cite{barack-review}).  In fact, simple scaling arguments (e.g., \cite{rosenthal}) suggest that even the leading self-force correction is not enough, and to achieve the desired accuracy one must keep terms \textit{second order} in the mass ratio.  While much theoretical and computational progress has been made on the first order problem, by contrast very little is known about second-order gravitational self-force.

Our previous work \cite{gralla-wald} (hereafter paper I) established a rigorous and systematic approach to the motion of small bodies in general relativity, based on a one-parameter-family of solutions to Einstein's equation.  Key elements are a far-zone limit (associated with viewing the body from far away) where the body shrinks to zero size and perturbs the external universe, and a near-zone limit (associated with viewing the body up close) where the body remains at fixed size and is perturbed by the external universe.  We developed the basic elements of the formalism to all orders in perturbation theory, but derived an equation of motion only to first order in the size/mass of the body.  The basic approach was to first compute the far-zone metric perturbation in some gauge, and then to seek a smooth gauge transformation such that the near-zone background metric becomes mass-centered (in that its mass dipole moment vanishes).  The value of the gauge vector on the background worldline then defines the perturbed position in the original gauge.  In paper I we derived an equation for the Lorenz gauge motion, while in a later paper \cite{gralla-gauge} (hereafter paper II) we derived an equation of motion holding in a larger class of gauges.

In the present work we identify a suitable notion of mass-centered at second order and define the second-order motion in an analogous way.  However, our derivation of an equation of motion proceeds in an entirely different manner.  One major change is that the approach is taken ``in reverse'': instead of beginning with an expression for a metric perturbation and seeking a gauge transformation to some mass-centered gauge, we instead begin with a series expansion for the general metric perturbation in a particular mass-centered gauge and consider the class of all smoothly related gauges.  This leads to a prescription (via an ``effective source'' method \cite{vega-detweiler,effective-source,barack-golbourn}) for computing the metric perturbation in such gauges, as well as a simple description of the motion in terms of the ``regular field'' employed in the effective source method.  In the present paper we assume for convenience that the spin and higher moments of the body are negligible, but there should be no obstacle of principle to relaxing these assumptions. 

A treatment of second-order gravitational self-force was given previously in  \cite{rosenthal}.  This approach is essentially axiomatic in that a number of properties that the force ``should'' have are assumed (principally, a list of ingredients from which it may be built\footnote{It is worth noting that one ingredient disallowed in \cite{rosenthal}, the angle-average, does appear in an expression for the force that holds in a larger class of gauges \cite{gralla-gauge}.}), and based on these assumptions a force expression is obtained.  By contrast, our approach is fundamental in that we begin with Einstein's equation for extended bodies in the limit of small size, and proceed by defining perturbed position and computing an equation it satisfies.  The approach of \cite{rosenthal} also contains a serious practical drawback in that it requires the first-order metric perturbation to be expressed in a gauge where the (first-order) self-force is zero.  Since a body will deviate secularly from its background motion as it loses energy to gravitational-wave emission, this gauge can only remain useful for a limited time and is inappropriate for calculations of inspiral (see discussion in section \ref{sec:inspiral}).  By contrast, our equation of motion holds in a class of gauges encompassing all possible motions.  Finally, the approach of \cite{rosenthal} encounters divergences of both the ``infrared'' (at spatial infinity) and ``ultraviolet'' (at the particle) varieties; while the ultraviolet divergences are regularized, the infrared divergences are left infinite.  By contrast, our derivation and result are well-defined.  A recent paper on second-order perturbation theory \cite{detweiler-2ndorder} also involves regularizations, and concludes with an equation whose mathematical legitimacy is unclear.\footnote{Equation (26) of \cite{detweiler-2ndorder} contains both delta functions and a term written as the second-order Einstein tensor acting on the distribution $h^{\textrm{1ret}}$.  Since products of distributions are not defined in general, it would require further analysis to give meaning to this term.  Since $h^{\textrm{1ret}}$ is sourced by a point particle delta function (within linearized theory), while point particle delta functions are not allowed in the full theory \cite{geroch-traschen}, it would be surprising if the second-order Einstein tensor of $h^{\textrm{1ret}}$ were a valid distribution.}

Very recently, Pound \cite{pound-2ndorder} has given an outline of a method to derive a prescription for computing the second order motion and metric perturbation of a small body.  The approach appears to contain many features similar to our own, although insufficient detail was given in \cite{pound-2ndorder} to enable a proper comparison of his approach to ours.

Our conventions are as follows.  We forgo the abstract index notation and work with coordinate components of tensors throughout.  We find this more convenient for discussing the perturbed position of the particle as well as for considering non-smooth coordinate transformations.  Greek indices label spacetime tensor components, while time and space components are denoted by $0$ and mid-alphabet Latin indices $i,j,\dots$, respectively.  Our sign conventions are those of Wald \cite{wald}.

\addtocounter{section}{1}

\section{Formalism and Outline}\label{sec:formalism}

The basic approach of paper I is to consider a one-parameter-family of spacetimes containing a body that shrinks to zero size and mass with the perturbation parameter, $\lambda$.  We build an appropriately shrinking body into the family by assuming the existence of a second limit that is designed to maintain any such body at fixed size (effectively ``zooming in'' on it).  More specifically, given a family of metrics $g_{\mu \nu}(\lambda)$ in coordinates ($t, x^i$), we introduce a scaled metric $\bar{g}_{\mu \nu} \equiv \lambda^{-2} g_{\mu \nu}$ and, for some time $t_0$, scaled coordinates $\bar{t} \equiv \lambda^{-1}(t-t_0)$ and $\bar{x}^i \equiv \lambda^{-1} x^i$.  Denoting the scaled metric in scaled coordinates by $\bar{g}_{\bar{\mu}\bar{\nu}}$, we then have the simple relationship
\begin{equation}\label{plugin}
\bar{g}_{\bar{\mu} \bar{\nu}}(\lambda; t_0; \bar{t},\bar{x}^i)
= g_{\mu \nu} (\lambda; t=t_0 + \lambda \bar{t}, x^i=\lambda \bar{x}^i),
\end{equation}
which relates components of the scaled metric in scaled coordinates to corresponding components of the original metric in the original coordinates.  One can construct perturbation series either off of the original metric or the scaled metric; we define $g^{(n)}_{\mu \nu} \equiv (1/n!) (\partial_\lambda)^n g_{\mu \nu}|_{\lambda=0}$ and $\bar{g}^{(n)}_{\bar{\mu} \bar{\nu}} \equiv (1/n!) (\partial_\lambda)^n \bar{g}_{\bar{\mu} \bar{\nu}}|_{\lambda=0}$, where derivatives are taken at fixed original and scaled coordinates, respectively.  These series are referred to as the far-zone and near-zone series, respectively.  

While we will always work with coordinate components in the original cartesian-like coordinates $(t,x^i)$, it is convenient to introduce spherical-like variables $r=\sqrt{\delta_{ij}x^i x^j}$ and $n^i=x^i/r$ (denoted $\vec{n}$ when representing a direction on the sphere).  The assumptions of paper I (adopted identically here) give the curve $r=\lambda=0$ (denoted $\gamma$) the interpretation of the lowest-order motion of the particle, and imply that $\gamma$ (assumed timelike) is in fact a geodesic.  This allows us to choose a coordinate system (such as Fermi normal coordinates) such that $g^{(0)}=\eta+O(r^2)$.  After making such a choice, our assumptions give the form of the far-zone series (defined only for $r>0$) to be
\begin{equation}\label{seriesfar}
\begin{array}{lllllllllll}
g^{(0)} & = \eta &+& 0 &+& a_{20} r^2 &+& a_{30} r^3 &+& O(r^4) \\
g^{(1)} & = a_{01} r^{-1} &+& a_{11} &+& a_{21} r &+& a_{31} r^2 &+& O(r^3) \\
g^{(2)} & = a_{02} r^{-2} &+& a_{12} r^{-1} &+& a_{22} &+& a_{32} r &+& O(r^2) \\
g^{(3)} & = a_{03} r^{-3} &+& a_{13} r^{-2} &+& a_{23} r^{-1} &+& a_{33} &+& O(r), \\
\end{array}
\end{equation}
where the $(a_{\mu \nu})_{nm}$ (tensor component indices spupressed above) are smooth functions of time and the two-sphere, $a_{nm} = a_{nm}(t,\vec{n})$.  Using equation \eqref{plugin}, one may obtain an expression for the near-zone series in terms of the $(a_{\mu \nu})_{nm}$,
\begin{align}\label{seriesnear}
\bar{g}^{(0)} & = \eta + a_{01}\bar{r}^{-1} + a_{02} \bar{r}^{-2} + a_{03} \bar{r}^{-3} + O(\bar{r}^{-4}) \nonumber \\
\bar{g}^{(1)} & = a_{11} + a_{12} \bar{r}^{-1} + a_{13} \bar{r}^{-2} + O(\bar{r}^{-3}) \nonumber \\  & \ + \bar{t} (\dot{a}_{01} \bar{r}^{-1} + \dot{a}_{02} \bar{r}^{-2} + \dot{a}_{03} \bar{r}^{-3} + O(\bar{r}^{-4}) ) \nonumber \\
\bar{g}^{(2)} & = a_{20} \bar{r}^2 + a_{21} \bar{r} + a_{22} + a_{23} \bar{r}^{-1} + O(\bar{r}^{-2}) \\ & \ + \bar{t} ( \dot{a}_{11} + \dot{a}_{12} \bar{r}^{-1} + \dot{a}_{13} \bar{r}^{-2} + O(\bar{r}^{-3}) ) \nonumber \\  & \ + \frac{1}{2} \bar{t}^2 ( \ddot{a}_{01} \bar{r}^{-1} + \ddot{a}_{02} \bar{r}^{-2} + \ddot{a}_{03} \bar{r}^{-3} + O(\bar{r}^{-4}) ) \nonumber \\
\bar{g}^{(3)} & = a_{30}\bar{r}^3 + a_{31}\bar{r}^2 + a_{32} \bar{r} + a_{33} + O(\bar{r}^{-1}) + \dots \nonumber,
\end{align}
where the $\bar{t}$-dependence of $\bar{g}^{(3)}$ (which goes up to $\bar{t}^3$) is left unexpressed.  In equation \eqref{seriesnear}, tensor indices (scaled on the LHS and unscaled on the RHS) have been suppressed, an overdot refers to a derivative with respect to the time argument of the $a_{nm}$, and the $a_{nm}$ are evaluated at $t=t_0$.  For example, in full notation the first line would read $\bar{g}^{(0)}_{\bar{\mu} \bar{\nu}} = \eta_{\mu \nu} + (a_{\mu \nu})_{01}(t_0,\vec{n}) \bar{r}^{-1} + \dots$.  

Notice that the ``columns'' of equation \eqref{seriesfar} correspond to near-zone perturbations in equation \eqref{seriesnear}.  For example, the near-zone background $\bar{g}^{(0)}$ corresponds to the first column in equation \eqref{seriesfar}, and the first near-zone perturbation $\bar{g}^{(1)}$ is specified by the second column (stationary part) and the time derivative of the first column (part linear in $\bar{t}$).  The alignment adopted in equation \eqref{seriesfar} is a helpful visualization tool for the relationship between the near-zone and far-zone perturbation series.

Equation \eqref{seriesnear} shows that near-zone background metric $\bar{g}^{(0)}$ is stationary and asymptotically flat.  Furthermore, the metric is written in adapted coordinates (the components are $\bar{t}$-independent and asymptotically Minkowskian), so that the mass dipole moment\footnote{By mass dipole moment of $\bar{g}^{(0)}$ we mean the $1/\bar{r}^2$, $\ell=1$ part of $(1/2) \bar{g}^{(0)}_{00}$.  Note that the mass dipole moment is contained in $a_{02}$, which is located at second order in the far-zone.} provides a measure of how ``off center'' the coordinates are.  In paper I we showed that a smooth first-order far-zone gauge transformation can always be made to eliminate the mass dipole moment of the near-zone background.  Since the new coordinates are then mass-centered, the new perturbed motion should vanish, suggesting that the old perturbed motion be defined to be the value of the gauge vector on the worldline.  This defines the first-order motion (in any allowed gauge\footnote{Since the metric perturbations are singular, non-smooth gauge transformations are allowed, the definition of motion in one gauge does not automatically define the motion in all other gauges.}) in terms of a far-zone gauge transformation to make the near-zone metric mass-centered at zeroth order.

We would similarly like to define the second order motion in terms of a far-zone gauge transformation that makes the near-zone metric mass-centered through first order.  However, our lowest-order notion of mass-centered (vanishing mass dipole) was sensible only because the metric components of $\bar{g}^{(0)}$ are always $\bar{t}$-independent and asymptotically Minkowskian.  It is clear from equation \eqref{seriesnear} that the perturbed metric $\bar{g}^{(0)}+\lambda \bar{g}^{(1)}$ will not necessarily satisfy these criteria.  However, if a gauge can be found where $\bar{g}^{(0)}+\lambda \bar{g}^{(1)}$ is in fact $\bar{t}$-independent and asymptotically Minkowskian and furthermore has no mass dipole, then we may regard this gauge as mass-centered.  We show below that such gauges do in fact exist, which defines the motion in these and smoothly related gauges.  However, unlike in the lower order case, we do not show that \textit{all} allowed gauges are smoothly related to a mass-centered gauge.  Instead, we simply find a mass-centered gauge and develop a prescription for working in any of the (large) class of smoothly related gauges.

The remainder of this paper is organized as follows.  In section \ref{sec:Pgauge}, we explicitly solve Einstein's equation in series in $r$ and $\lambda$ to determine the general solution compatible with our assumptions (to the relevant orders in $r$ and $\lambda$) up to coordinate freedom.  In particular, this establishes a convenient local inertial coordinate system for the far-zone background metric (named RWZ coordinates), and a convenient mass-centered gauge choice\footnote{We refer to finite-$\lambda$ coordinate transformations that preserve the metric components of the background metric as ``gauge transformations''.} (named P gauge) for the metric perturbations.  We use our P gauge solution in two important ways.  First, in section \ref{sec:Psmooth} we use the explicit singular form of the $P$ gauge solution to identify appropriate ``singular fields'' for use in an effective source prescription for computing the global metric perturbation in smoothly related gauges ($P$-smooth gauges).  Second, in section \ref{sec:motion} we use the mass-centered property of the $P$ gauge to deduce a simple prescription for determining the motion in $P$-smooth gauges.  We do not ascribe any fundamental status to our particular class of $P$-smooth gauges, and in section \ref{sec:choiceP} we discuss how the paper could have proceeded (identically) if an alternative class of gauges were used.  In section \ref{sec:inspiral} we discuss incorporating our results into a formalism for long-term waveform generation.  Finally in section \ref{sec:summary} we summarize our prescription for computing the second-order motion and metric perturbation of a small body.

\section{Local Metric in P-gauge}\label{sec:Pgauge}

We now explicitly construct a gauge that is mass centered in the sense discussed in section \ref{sec:formalism}, i.e., a gauge in which the near-zone metric is $\bar{t}$-independent and asymptotically Minkowskiian and through first order in $\lambda$.  We will call this gauge the P gauge, where the P stands for ``particular'', in order to emphasize that other mass-centered gauges could have been chosen.  (We discuss this freedom in more detail in section \ref{sec:choiceP}.) We perform our computations using the near-zone perturbation series.  While it is necessary to proceed only to first order in $\lambda$ to establish that a gauge is mass-centered, the mass-centered coordinate choice influences the form of terms at higher order in near-zone perturbation theory, many of which will be needed for the later analysis of the paper.  In performing our near-zone calculations, we will in fact have to proceed through \textit{third} order in $\lambda$.

We begin our computations with the background near-zone metric.  Since this metric is stationary and asymptotically flat, it is characterized by multipole moments \cite{multipole-moments}.  We treat a body with negligible spin and higher moments, and therefore take the spin and higher moments of this metric to vanish.  Thus the near-zone background metric is simply the Schwarzschild (exterior) metric\footnote{While for convenience we will make statements about an ``entire'' metric, it should be borne in mind that we only require that such statements hold to the orders explicitly displayed in the paper.  (These orders have been carefully chosen for consistency with all statements made.)  For example, in this case we say that the metric is Schwarzschild, but in fact we only require it to match Schwarzschild to $O(\bar{r}^{-3})$ (see equation \eqref{seriesnear}).  Thus we in fact only assume that the spin and quadrupole moments vanish---the effects of higher moments are automatically negligible at the present level of approximation (that is, these moments do not appear at the orders in $\lambda$ pursued in this paper).} for all time $t_0$.  While in principle the mass may depend on time $t_0$, in paper I it was shown to be constant.  We label the mass by $M$ and choose Cartesian isotropic coordinates for $\bar{g}^{(0)}$ (at all time $t_0$).

Since the metric components of $\bar{g}^{(0)}$ are then independent of $t_0$,  by \eqref{seriesnear} the near-zone perturbation $\bar{g}^{(1)}$ must be independent of time $\bar{t}$.  Furthermore, equation \eqref{seriesnear} shows that the perturbation is asymptotically constant.  Standard Schwarzschild perturbation results \cite{regge-wheeler,zerilli} then imply that its only physical effect can be to perturb the multipole moments of $\bar{g}^{(0)}$.  In line with our choice to consider a body with no spin and higher moments, we set the perturbed spin and higher moments to zero.  While in principle there could be a perturbation to the mass, it was shown in paper II that the perturbed mass does not evolve with time.  Therefore this quantity may as well be ``renormalized'' into the background mass, or equally acceptably simply set to zero.   We will set the perturbed mass (and higher moments) to zero.  With these physical choices the first near-zone perturbation is pure gauge, and we choose it to vanish.  

With the above choices the near-zone metric agrees with the ordinary (mass-centered) Schwarzschild metric in Cartesian isotropic coordinates through first order in $\lambda$, and therefore fits our notion of being mass-centered.  Explicitly, the perturbation series \eqref{seriesnear} is now given by
\begin{align}
\bar{g}^{(0)} & = \eta + \mathcal{M}^{(1)}\bar{r}^{-1} + \mathcal{M}^{(2)} \bar{r}^{-2} + \mathcal{M}^{(3)} \bar{r}^{-3} + O(\bar{r}^{-4}) \label{gbar0} \\
\bar{g}^{(1)} & = O(\bar{r}^{-3}) + \bar{t} O(\bar{r}^{-4}) \label{gbar1} \\
\bar{g}^{(2)} & = a_{20} \bar{r}^2 + a_{21} \bar{r} + a_{22} + O(\bar{r}^{-1}) \nonumber \\ & \qquad + \bar{t} O(\bar{r}^{-3}) + \bar{t}^2 O(\bar{r}^{-4}) \label{gbar2} \\
\bar{g}^{(3)} & = a_{30}\bar{r}^3 + a_{31}\bar{r}^2 + a_{32} \bar{r} + O(\bar{r}^0)) \nonumber \\ & + \bar{t} (\dot{a}_{20} \bar{r}^2 + \dot{a}_{21} \bar{r} + \dot{a}_{22} + O(\bar{r}^{-1}) ) \nonumber \\ & + \bar{t}^2 O(\bar{r}^{-3})  + \bar{t}^3 O(\bar{r}^{-4}) \label{gbar3},
\end{align}
where $\mathcal{M}^{(n)}$ is the $n^{\textrm{th}}$-order term of the Schwarzschild metric in Cartesian isotropic coordinates.
\begin{align}
\mathcal{M}^{(1)}_{\mu \nu} & = 2M \left( \eta_{\mu \nu}+2t_\mu t_\nu \right) \label{M1} \\
\mathcal{M}^{(2)}_{\mu \nu} & = M^2 \left( \frac{3}{2} \eta_{\mu \nu}-\frac{1}{2}t_\mu t_\nu \right) \label{M2} \\
\mathcal{M}^{(3)}_{\mu \nu} & = M^3 \left( \frac{1}{2} \eta_{\mu \nu}+2t_\mu t_\nu \right). \label{M3}
\end{align}
with $t_\mu=(-1,0,0,0)$.

 We now consider the second-order near-zone metric perturbation, $\bar{g}^{(2)}$.  Since the first near-zone perturbation vanishes, the second perturbation satisfies the linearized Einstein equation off of the Schwarzschild background.  Equation \eqref{gbar2} shows that our perturbation is $\bar{t}$-independent to the relevant order, so that we may restrict consideration to stationary solutions.  We use the Regge-Wheeler-Zerilli (RWZ) formalism \cite{regge-wheeler,zerilli}, where one decomposes the perturbation into a sum of (tensor) spherical harmonic modes labeled by azimuthal number $\ell$.  For $\ell=0$ and $\ell=1$ modes, the general stationary solution (up to gauge) has $\bar{r}\rightarrow\infty$ asymptotic behavior of $\bar{r}^{-(\ell+1)}$, while for $\ell>1$ the general stationary solution (up to gauge) is given by a linear combination of functions behaving as $\bar{r}^{-(\ell+1)}$ and $\bar{r}^{\ell}$.  From these considerations, comparison with equation \eqref{gbar2} shows that the general solution (up to gauge) for $\bar{g}^{(2)}$ of our form is pure $\ell=2$ to the displayed orders.  This solution is characterized by ten constants (one for each $m$-mode of each parity), which are conveniently represented as two constant symmetric trace-free (STF) tensors in three-dimensional Euclidean space.  (Excellent reviews of the STF approach to spherical harmonic decompositions are found in \cite{blanchet-damour-review, thorne-review}.)  In our computations we employ the RWZ formalism as presented in \cite{martel,martel-poisson,hopper-evans}, translating the results into Cartesian isotropic coordinates and STF language.  We use the closed-form expressions for the stationary master functions given in \cite{field-hestaven-lau}.  We find that the general solution (in Regge-Wheeler gauge) for our $\bar{g}^{(2)}$ may be written in terms of two arbitrary STF tensors $\mathcal{E}_{ij}$ and $\mathcal{B}_{ij}$ by
\begin{align}
\bar{g}^{(2)}_{00} & = \mathcal{E}_{ij}n^i n^j( -\bar{r}^2 + 2M \bar{r} + \frac{3}{2} M^2 ) + O(\bar{r}^{-1}) \label{gbar2-00}\\
\bar{g}^{(2)}_{i0} & = \epsilon_{ijk} n^j \mathcal{B}^{k}_{\ l} n^l (\frac{2}{3} \bar{r}^2 + \frac{2}{3} M \bar{r} - \frac{1}{6} M^2 ) + O(\bar{r}^{-1}) \label{gbar2-i0} \\
\bar{g}^{(2)}_{ij} & = \mathcal{E}_{kl}n^k n^l\big[ \delta_{ij} (-\bar{r}^2 -4 M \bar{r} - 5M^2) \nonumber \\ & \qquad \qquad \ \ \ + n_i n_j (2 M \bar{r} - 4M^2) \big] + O(\bar{r}^{-1}), \label{gbar2-ij}
\end{align}
where $\epsilon_{ijk}$ is the Cartesian Levi-Civita symbol.  The STF tensors $\mathcal{E}_{ij}$ and $\mathcal{B}_{ij}$ may depend on the time $t_0$ at which the near-zone limit is taken, but are independent of the near-zone background  coordinates $\bar{t},\bar{x}^i$.  Equations \eqref{gbar2-00}-\eqref{gbar2-ij} determine the unknown functions $a_{20}$,$a_{21}$,$a_{22}$ in a particular gauge.

We now turn to the third order near-zone perturbation, $\bar{g}^{(3)}$.  This perturbation also satisfies the linearized Einstein equation off of the Schwarzschild background (on account of the vanishing of the first perturbation).  We use the RWZ formalism to find the general solution for $\bar{g}^{(3)}$ consistent with equation \eqref{gbar3}.  From general considerations of the sort discussed for the second-order perturbation, above, this solution contains only $\ell=2$ and $\ell=3$ modes (up to gauge).  The $\bar{t}$-dependence of the perturbation is fixed entirely by $\bar{g}^{(2)}$ (see equation \eqref{gbar3}).  For the part independent of $\bar{t}$, the RWZ formalism yields
\begin{align}
\bar{g}^{(3)}_{00}|_{\bar{t}=0} & = \mathcal{E}_{ijk}n^i n^jn^k( -\frac{1}{3}\bar{r}^3 + \frac{2}{3}M \bar{r}^2 + \frac{7}{12} M^2 \bar{r} ) \nonumber \\ & + \delta \mathcal{E}_{ij}n^i n^j( -\bar{r}^2 + 2M \bar{r} ) + O(\bar{r}^{0}) \label{gbar3-00}\\
\bar{g}^{(3)}_{i0}|_{\bar{t}=0} & = \epsilon_{ijk} n^j \mathcal{B}^{k}_{\ l m} n^l n^m (\frac{2}{3} \bar{r}^3 + \frac{4}{9} M \bar{r}^2 - \frac{2}{9} M^2 \bar{r} ) \nonumber \\ & + n_i \dot{\mathcal{E}}_{kl} n^k n^l(-\frac{2}{3} \bar{r}^3 - \frac{7}{3} M \bar{r}^2 - \frac{5}{3} M^2 \bar{r}) \nonumber \\ & + \epsilon_{ijk} n^j \delta \mathcal{B}^{k}_{\ l} n^l (\frac{2}{3} \bar{r}^2 + \frac{2}{3} M \bar{r} ) + O(\bar{r}^{0}) \label{gbar3-i0}\\
\bar{g}^{(3)}_{ij}|_{\bar{t}=0} & = \mathcal{E}_{klm}n^k n^l n^m\big[ \delta_{ij} (-\frac{1}{3}\bar{r}^3 - M \bar{r}^2 - \frac{13}{12}M^2\bar{r}) \nonumber \\ & \qquad \qquad \qquad \qquad \ \ \ + n_i n_j (\frac{1}{3} M \bar{r}^2 - \frac{2}{3} M^2\bar{r}) \big] \nonumber \\ & + n_{(i}\epsilon_{j)kl}n^k \dot{\mathcal{B}}^l_{\ m} n^m(\frac{1}{3}\bar{r}^3 + 2 M \bar{r}^2 - \frac{67}{12}M^2\bar{r}) \nonumber \\ & + \delta \mathcal{E}_{kl}n^k n^l\big[ \delta_{ij} (-\bar{r}^2 -4 M \bar{r}) +  2 M \bar{r} n_i n_j \big] \nonumber \\ & \qquad \qquad + O(\bar{r}^{0}),\label{gbar3-ij}
\end{align}
where we have introduced STF tensors $\mathcal{E}_{ijk}$, $\mathcal{B}_{ijk}$, $\delta \mathcal{E}_{ij}$, and $\delta \mathcal{B}_{ij}$, and the overdot indicates a derivative with respect to $t_0$.  Equations \eqref{gbar3-00}-\eqref{gbar3-ij} determine the unknown functions $a_{30}$,$a_{31}$,$a_{32}$ in a mass-centered gauge.  We have therefore constructed the perturbation series in a particular gauge to the relevant orders.

We note that the strategy of using the RWZ formalism in the near-zone has been employed before, most notably in \cite{detweiler-1storder,detweiler-review}.  The main difference (besides the different coordinate choices) is that \cite{detweiler-1storder,detweiler-review} impose boudary conditions of regularity at the horizon of a black hole.  By contrast, we treat an arbitrary body and impose boundary conditions only at infinity.  Indeed, it is a result of our computations that no interior boundary conditions are necessary to fix the form at the relevant orders, which corresponds (after the analysis of the paper) to the result that the motion of the body is independent of its detailed composition at the perturbative orders considered.

\subsection{Far Zone Expressions}

Our formulae for the near-zone metric determine the displayed coefficients $a_{nm}$ in equation \eqref{seriesnear}.  These coefficients may then be used to reconstruct the far-zone series via equation \eqref{seriesfar}.  This gives the general far-zone solution to the Einstein equation to the relevant orders in $r$ and $\lambda$ in a particular gauge.  However, rather than using this gauge as our P gauge, we instead first make a particular first-order (far-zone) gauge transformation, which is designed to make the first-order metric perturbation satisfy the Lorenz condition, while preserving the mass-centered property.  Using this gauge as our P gauge ensures that the P-smooth class includes Lorenz gauges, which facilitates comparison with previous work using Lorenz gauge, as well as allows Lorenz-gauge numerical results to be incorporated into second-order calculations based on our prescription.  However, we emphasize that (unlike in some previous work) the Lorenz gauge plays no fundamental role in our analysis.  Our choice to perform an additional Lorenz-motivated gauge transformation (i.e., our Lorenz-motivated choice of P gauge) affects the details of many complicated formulae throughout the paper, but has otherwise no influence on our prescription.  We specifically discuss alternative choices of P gauge in section \ref{sec:choiceP}.

\textit{0. Notation} In the remainder of the paper we will refer to the far-zone background, first perturbation, second perturbation, and third perturbation as $g$, $h$, $j$, and $k$, respectively.  (That is, $g=g^{(0)}$, $h=g^{(1)}$ and $j=g^{(2)}$, $k=g^{(3)}$.)  This facilitates the introduction of many necessary new superscripts and other modifiers.  In order to avoid any potential ambiguity, the one-parameter-family of metrics will always be referred to with the $\lambda$-dependence indicated, $g(\lambda)$.

\textit{1. Background.}  Reconstructing the background metric $g$ from the near-zone solution gives
\begin{equation}\label{gform}
g = \eta + a_{20}(t) r^2 + a_{30}(t) r^3 + O(r^4),
\end{equation}
with
\begin{align}
(a_{00})_{20} & = - \mathcal{E}_{ij} n^i n^j \\
(a_{i0})_{20} & = \frac{2}{3} \epsilon_{ijk} n^j \mathcal{B}^k_{\ l} n^l \\
(a_{ij})_{20} & = - \delta_{ij} \mathcal{E}_{kl}n^k n^l \\
\nonumber \\
(a_{00})_{30} & = -\frac{1}{3} \mathcal{E}_{ijk} n^i n^j n^k \\
(a_{i0})_{30} & = \frac{2}{3} \epsilon_{ijk} n^j \mathcal{B}^k_{\ lm} n^l n^m - \frac{2}{3} n_i \dot{\mathcal{E}}_{kl} n^k n^l \\
(a_{ij})_{30} & = - \frac{1}{3} \delta_{ij} \mathcal{E}_{klm}n^k n^l n^m+ \frac{1}{3} n_{(i} \epsilon_{j)kl}n^k \dot{\mathcal{B}}^l_{\ m} n^m.
\end{align}
We may now interpret the STF tensors $\mathcal{E}_{ij}$, $\mathcal{B}_{ij}$, $\mathcal{E}_{ijk}$ and $\mathcal{B}_{ijk}$ by computing the Riemann tensor of $g$.  It is then straightforward to see that
\begin{align}
\mathcal{E}_{ij} & = R_{0i0j}|_{\gamma} \label{Eij} \\
\mathcal{B}_{ij} & = -\frac{1}{2} \epsilon^{kl}_{\ \ i}R_{0jkl}|_{\gamma} \label{Bij} \\
\mathcal{E}_{ijk} & = \nabla_{(k} R_{|0|i|0|j)}|_{\gamma} \label{Eijk} \\
\mathcal{B}_{ijk} & = \frac{3}{16} \epsilon^{lm}_{\ \ (i}\nabla_j R_{|0|k)lm}|_{\gamma}. \label{Bijk}
\end{align}
Thus our coordinate system for the background metric expresses an arbitrary vacuum metric in terms of the curvature on an arbitrary timelike geodesic $x^i=0$.  We refer to these coordinates as RWZ coordinates, after the use of the Regge-Wheeler-Zerilli gauge in solving the near-zone equations.  Our metric agrees with the Fermi normal coordinate metric (e.g., \cite{poisson-review}) to $O(r)$ and with the Thorne-Hartle-Zhang metric \cite{THZ,poisson-review,detweiler-review} to $O(r^2)$.  The form of the metric at $O(r^3)$ appears to be new.

\textit{2. First Perturbation.} Reconstructing the first far-zone perturbation gives a specific expression for $h$ in terms of $\mathcal{E}_{ij}$, $\mathcal{B}_{ij}$, $\mathcal{E}_{ijk}$, $\mathcal{B}_{ijk}$, $\delta \mathcal{E}_{ij}$, and $\delta \mathcal{B}_{ij}$.  Instead of adopting this expression as our $P$ gauge, however, we first make a particular gauge transformation generated by the gauge vector
\begin{align}
v_0 & = -\frac{10}{9} r^3 \dot{\mathcal{E}}_{ij} n^i n^j \label{v0} \\
v_i & = r^2 \left( 2 \mathcal{E}_{ij} n^j - n_i \mathcal{E}_{jk} n^j n^k \right) \nonumber \\ & + r^3 \left( \frac{1}{2} \mathcal{E}_{ijk} n^j n^k - \frac{1}{6} n_i \mathcal{E}_{klm}n^k n^l n^m - \frac{2}{3} \epsilon_{ijk} n^k n^l \dot{\mathcal{B}}^j_{\ l} \right). \label{vi}
\end{align}
This gauge transformation is designed to make $h^P$ satisfy the Lorenz condition, as may be checked by direct computation using the formulae below.  As will be displayed explicitly in equation \eqref{seriesfar2}, below, the transformation does not affect the mass-centered property of the coordinates. After performing the transformation we denote the resulting perturbation by $h^P$, which is given by
\begin{equation}\label{hP}
h^P = \mathcal{M}^{(1)}r^{-1} + a_{21} r + a_{31} r^2 + O(r^3)
\end{equation}
with
\begin{align}
(a_{00})_{21} & = 2 M \mathcal{E}_{ij} n^i n^j \label{a2100} \\
(a_{i0})_{21} & = \frac{2}{3} M \epsilon_{ijk} \mathcal{B}^k_{\ l} n^l \label{a21i0} \\
(a_{ij})_{21} & = - 2 M ( \delta_{ij} \mathcal{E}_{kl}n^k n^l + 2 \mathcal{E}_{ij}) \label{a21ij}
\end{align}
and
\begin{align}
(a_{\mu \nu})_{31} & = (a^S_{\mu \nu})_{31} + (a^H_{\mu \nu})_{31} \\
\nonumber \\
(a^S_{00})_{31} & = \frac{2}{3} M \mathcal{E}_{ijk} n^i n^j n^k \label{aS00} \\
(a^S_{i0})_{31} & = \frac{4}{9} M \epsilon_{ijk} n^j \mathcal{B}^k_{\ lm} n^l n^m \nonumber \\ & - \frac{2}{9} M ( \dot{\mathcal{E}}_{ij}n^j - n_i \dot{\mathcal{E}}_{kl}n^k n^l ) \label{aSi0} \\
(a^S_{ij})_{31} & = M \left( -\frac{2}{3} \delta_{ij} \mathcal{E}_{klm} n^k n^l n^m - 2 \mathcal{E}_{ijk}n^k \right) \nonumber \\ & + \frac{2}{3} M \left( n_{(i} \epsilon_{j)kl}n^k \dot{\mathcal{B}}^l_{\ m}n^m - 2 \dot{\mathcal{B}}_{(i}^{\ \ l} \epsilon_{j)kl} n^k \right) \label{aSij} \\
\nonumber \\
(a_{00}^H)_{31} & = - \delta \mathcal{E}_{ij}n^i n^j \label{aH00} \\
(a_{i0}^H)_{31} & = \frac{2}{3} \epsilon_{ijk} n^j \delta \mathcal{B}^k_{\ l} n^l \label{aHi0} \\
(a_{ij}^H)_{31} & = - \delta_{ij} \delta \mathcal{E}_{kl}n^k n^l. \label{aHij}
\end{align}
We have split $a_{31}$ into S and H pieces in order to make a similar split $h^P = h^S + h^H+O(r^3)$,
\begin{align}
h^S_{\mu \nu} & = \mathcal{M}^{(1)}_{\mu \nu} r^{-1} + (a_{\mu \nu})_{21} r + (a^S_{\mu \nu})_{31} r^2 + O(r^3) \label{hS} \\
h^H_{\mu \nu} & = (a^H_{\mu \nu})_{31} r^2 + O(r^3). \label{hH}
\end{align}
The reason for this split will become clear when the ``singular field'' $h^S$ is employed in the following section as part of a prescription for computing the metric perturbation.  The guiding principle is that $h^S$ be determined by the background metric (containing only $\mathcal{E}_{ij}, \mathcal{B}_{ij}, \mathcal{E}_{ijk}, \mathcal{B}_{ijk}$, and not the unknown $\delta \mathcal{E}_{ij}$ and $\delta \mathcal{B}_{ij}$) and that the remainder $h^H$ be $C^2$.  There are many other choices besides ours that satisfy these properties, and we could equally well have made these choices.  Our choices also have the additional properties that $h^S$ and $h^H$ separately solve the field equations to the displayed orders.\footnote{This is most easily seen by checking that $h^H$ is a solution, a computation that requires only the leading order term  $g=\eta$ of the background.  Since the sum $h^P=h^H+h^S$ is by construction a solution for $r>0$, it follows that $h^S$ is also a solution for $r>0$.}  For future use, we relate $\delta \mathcal{E}_{ij}$ and $\delta \mathcal{B}_{ij}$ to $h^H$ by computing the linearized Riemann tensor of $h^H$, finding (cf. equations \eqref{Eij} and \eqref{Bij})
\begin{align}
\delta \mathcal{E}_{ij} & = \left. R^{(1)}_{0 i 0 j}[h^H] \right|_{\gamma} \label{deltaE} \\
\delta \mathcal{B}_{ij} & = -\frac{1}{2} \epsilon^{kl}_{\ \ i} \left.  R^{(1)}_{0 j k l}[h^H] \right|_{\gamma}. \label{deltaB}
\end{align}
Here $R^{(1)}_{\mu \nu \rho \sigma}[h]$ is defined as a function of a symmetric rank 2 tensor $h_{\mu \nu}$ by $R^{(1)}_{\mu \nu \rho \sigma}[\partial_\lambda g|_{\lambda=0}]=\partial_\lambda R_{\mu \nu \rho \sigma}(\lambda)|_{\lambda=0}$ for a smooth one-parameter-family $g(\lambda)$.

\textit{3. Second Perturbation.} We reconstruct $j$ from the near-zone solution and take into account the effects on $j$ of the first-order gauge transformation, equations \eqref{v0}-\eqref{vi}.  The second-order perturbation is given by
\begin{equation}\label{jP}
j^P = \mathcal{M}^{(2)}r^{-2} + a_{22} + a_{32} r + O(r^2)
\end{equation}
with
\begin{align}
(a_{00})_{22} & = -3 M^2 \mathcal{E}_{ij} n^i n^j  \label{a2200} \\
(a_{i0})_{22} & =  \frac{10}{3} M^2 \epsilon_{ijk} n^j n^l \mathcal{B}^k_{\ l}  \label{a22i0} \\
(a_{ij})_{22} & =  8 M^2 \mathcal{E}_{ij} - M^2 \mathcal{E}_{kl}n^k n^l \left( 6\delta_{ij} + n_i n_j \right) \label{a22ij} \\
(a_{00})_{32} & = -\frac{5}{3} \mathcal{E}_{ijk} n^i n^j n^k  + \frac{3}{2} M \delta \mathcal{E}_{ij}n^i n^j \label{a3200} \\
(a_{i0})_{32} & = \frac{2}{3} \epsilon_{ijk} n^j \mathcal{B}^k_{\ lm} n^l n^m - \frac{2}{3} n_i \dot{\mathcal{E}}_{kl} n^k n^l \nonumber \\ & - \frac{1}{6} M \epsilon_{ijk} n^j \delta \mathcal{B}^k_{\ l} n^l \label{a32i0} \\
(a_{ij})_{32} & = - \frac{1}{3} \delta_{ij} \mathcal{E}_{klm}n^k n^l n^m+ \frac{1}{3} n_{(i} \epsilon_{j)kl}n^k \dot{\mathcal{B}}^l_{\ m} n^m \nonumber \\& - M \delta_{ij} \delta \mathcal{E}_{kl}n^k n^l. \label{a32ij}
\end{align}

Note that we have not made a second-order gauge transformation, analogous to the transformation \eqref{v0}-\eqref{vi} made at first order.  The first-order gauge transformation was designed to make $h^P$ satisfy the Lorenz condition, which was desirable because of the long history of use of Lorenz gauge in both theoretical and computational work at first order.  For second-order perturbation theory, the relevant previous work is \cite{pound,poisson-review,pound-2ndorder}, where the Lorenz condition was imposed on the second-order metric perturbation.\footnote{Note that when specialized to a flat background spacetime, this differs from the harmonic gauge condition used in Post-Newtonian theory by terms involving the first-order perturbation.}  In the interest of comparison, we have investigated whether this condition may be imposed within our formalism.  We have found that it appears necessary to introduce $r \log r$ terms in to the metric perturbation in order to impose this condition.  This directly violates the metric form required by our assumptions (equation \eqref{seriesfar}), and, if allowed, would lead to $\lambda \log \lambda$ terms in the near-zone series by equation \eqref{plugin}.   Since a smooth near-zone perturbation series is an essential ingredient in our justification (see paper I) of the relevance of our perturbation series to small (but extended) bodies, we take the viewpoint that such a far-zone gauge is too singular to sensibly describe a small body, at least within our current approach.\footnote{The appearance of $\log$ terms at second order in the Lorenz gauge was also found in the gauge-relaxed formalism of \cite{pound}.  This has an analogous singular effect on the near-zone metric; this effect is not discussed.}

\textit{4. Third Perturbation.} Reconstructing the third-order metric perturbation from the near-zone yields
\begin{equation}
k^P = \mathcal{M}^{(3)} r^{-3} + O(r^{-1}).
\end{equation}
 While it would have been straightforward to compute the terms proportional to $r^{-1}$ and $r^{0}$ (i.e., $a_{23}$ and $a_{33}$) from our near-zone expression (plus the effects of the first order gauge transformation, equations \eqref{v0}-\eqref{vi}), these terms are not relevant for our analysis.  Note that the first-order gauge transformation, equations \eqref{v0}-\eqref{vi}, has had no effect on the displayed orders.  We now collect the results of this section in the form of equation \eqref{seriesfar},
\begin{equation}\label{seriesfar2}
\begin{array}{lllllllllll}
g & = \eta &+& 0 &+& a_{20} r^2 &+& a_{30} r^3 &+& O(r^4) \\
h^P & = \mathcal{M}^{(1)} r^{-1} &+& 0 &+& a_{21} r &+& a_{31} r^2 &+& O(r^3) \\
j^P & = \mathcal{M}^{(2)} r^{-2} &+& 0 &+& a_{22} &+& a_{32} r &+& O(r^2) \\
k^P & = \mathcal{M}^{(3)} r^{-3} &+& 0 &+& O(r^{-1}), &\ & \ &\ & \ 
\end{array}
\end{equation}
where the $a_{nm}$ are now given explicitly by the formulae in this section.  In this form it is easily seen that the near-zone background (first column) is Schwarzschild and the first near-zone perturbation (second column) vanishes, so that the $P$ gauge is indeed mass-centered.

\textit{5. Summary of Results.} Equation \eqref{seriesfar2}, together with the preceding expressions for the $a_{nm}$, is the main result of this section.   This expression provides a series expansion in $r$ for general zeroth, first, second, and third-order metric perturbation subject to our assumptions, expressed in a particular mass-centered gauge, known as $P$-gauge.  For use in the following section, we have also isolated off a particular singular portion of $h^P$, denoted $h^S$.  We have used the tensor analysis package \textit{xTensor} \cite{xTensor} for the software package \textit{Mathematica} \cite{mathematica} to perform many of the computations in this section.  We have verified by direct computation (taking several hours on a personal computer) that the metric $g + \lambda h^P + \lambda^2 j^P + \lambda^3 k^P$ satisfies Einstein's equation to the relevant orders in $\lambda$ and $r$.

\section{Global Metric in $P$-smooth gauges}\label{sec:Psmooth}

In the previous section the general solution for the metric $g(\lambda)$ was determined in series in $r$ and $\lambda$, subject to particular coordinate choices.  Since the motion is also known in these coordinates (it is given by the coordinates of the background geodesic $\gamma$), we have at some level determined the general solution to our problem.  Of course, this general solution is of no use in practice, since it contains undetermined parameters (with no physical interpretation) and gives the metric only locally near $r=0$.  Nevertheless, this analysis has revealed the structure of the general solution near $r=0$, which will allow us to develop a prescription for obtaining the global metric perturbation in a P-smooth gauge in situations of physical interest, as described below.

Given our assumptions on the one-parameter-family, Einstein's equation implies in the far-zone that
\begin{align}
G^{(1)}_{\mu \nu}[h] 
& = 0 \qquad (\textrm{for } r>0) \label{G1} \\
G^{(1)}_{\mu \nu}[j] + G^{(2)}_{\mu \nu}[h] & = 0 \qquad (\textrm{for } r>0), \label{G2}
\end{align}
where $G^{(1)}$ and $G^{(2)}$ are the first and second order Einstein operators, respectively.  When combined with the assumed form of the metric perturbations near $r=0$ (equation \eqref{seriesfar}), these equations provide the complete description required to compute $h$ and $j$ in a given situation of interest (i.e., once suitable initial and/or boundary conditions have been prescribed).  In practice, however, it may be difficult to ensure that a numerical solution have the correct divergent behavior near $r=0$.  Furthermore, it is far from obvious how to ensure that the metric perturbation will be determined in a gauge for which we define the motion.

A solution to both of these problems is to use our knowledge of the general P-gauge series solution near $r=0$ to ``regularize'' the differential equation.  One simply subtracts off the known singular behavior and evolves the regular remainder.  This type of numerical technique was introduced into the field of self-force computation by \cite{barack-golbourn, vega-detweiler}, and is now generally known as the ``effective source approach'' \cite{effective-source}.  At first order, our approach is equivalent to the standard approach, except that we are not restricted to the Lorenz gauge, and instead allow the use of any gauge condition that gives rise to a sufficiently regular ``regular field''.  Our presentation of the method differs in that we do not make use of $\delta$-function sources, instead working directly with our assumed form of the metric perturbation for $r>0$.\footnote{In paper I, we proved that our assumptions in fact \textit{imply} a delta-function source for $h$ (regarded as a distribution) at first order.  From the point of view of developing an effective source description from our assumptions, such a delta-function description would appear only as an unnecessary intermediary.}

\subsection{First Order}\label{sec:firstorder}

In the previous section we constructed the general solution for the first-order metric perturbation in series in $r$ in a particular gauge.  We refer to this gauge as the P gauge and denote the perturbation by $h^P$.  In a general smoothly related gauge, the metric perturbation is given by
\begin{align}
h & = h^P - \mathcal{L}_\xi g + O(r^3) \label{hPBG} \\
  & = h^S + h^H - \mathcal{L}_\xi g + O(r^3), \label{hSHBG}
\end{align}
where the split of $h^P$ into $h^S$ and $h^H$ was introduced in equation \eqref{hS}.\footnote{The error terms in equations \eqref{hPBG} and \eqref{hSHBG} are redundant with those in the definitions of $h^P$, $h^S$ and $h^H$, but we include the error terms as a reminder of the local nature of $h^P$, $h^S$ and $h^H$.} Recall that $h^S=O(1/r)$ is a singular approximate solution to the linearized Einstein equation specified by the background curvature tensors $\mathcal{E}_{ij},\mathcal{B}_{ij},\mathcal{E}_{ijk},\mathcal{B}_{ijk}$, while $h^H=O(r^2)$ is a $C^2$ approximate solution given in terms of undetermined parameters $\delta \mathcal{E}_{ij}$ and $\delta \mathcal{B}_{ij}$, which encode its Riemann curvature via equations \eqref{deltaE} and \eqref{deltaB}.  The sum, $h^P$, represents the general solution with a particular gauge choice, up to $O(r^3)$ errors.   We emphasize that our $h^P$, $h^S$, and $h^H$ are given only as approximate solutions near $r=0$; none of these quantities has a finite-$r$ definition from which the series expansions emerge.  This is in contrast to the singular field of \cite{detweiler-whiting}, which is defined in a normal neighborhood through the use of Hadamard Green's function techniques in the Lorenz gauge.  We not checked if our $h^S$ agrees with a series expansion of the Detweiler-Whiting singular field.\footnote{In \cite{detweiler-review} a singular field was constructed in a manner similar to ours (but using different coordinate choices and notation).  It was then claimed that this singular field agrees with the Detweiler-Whiting singular field up to errors of $O(r^2)$.  It seems likely that our singular field agrees with that of \cite{detweiler-review} (and therefore with the Detweiler-Whiting singular field) at this order.}

Implementing the effective source approach requires choosing an (arbitrary) extension of $h^S$ to the entire manifold (minus  $r=0$).  We will distinguish extended quantities with a ``hat'': Let $\hat{h}^S$ denote an arbitrary function on the manifold (minus $r=0$) such that $\hat{h}^S$ agrees with $h^S$ (equation \eqref{hS}) to all displayed orders in $r$ (i.e., to $O(r^2)$).  We then define a global ``regular field'' $\hat{h}^R$ in terms of the metric perturbation $h$ by
\begin{align}
\hat{h}^R & = h - \hat{h}^S \label{hRext} \\
          & = h^H - \mathcal{L}_\xi g + O(r^3). \label{hRext2}
\end{align}
Plugging equation \eqref{hRext} into the linearized Einstein equation \eqref{G1} gives
\begin{equation}
G^{(1)}[\hat{h}^R] = -G^{(1)}[\hat{h}^S] \qquad (\textrm{for } r>0).
\end{equation}
By construction we have that $G^{(1)}[\hat{h}^S]$ is $O(r)$, so that the right-hand side is in fact $O(r)$.   Thus the ``source'' $-G^{(1)}[\hat{h}^S]$ is $C^0$, and we may in fact drop the requirement that $r>0$.  We may then write the first-order equation as simply
\begin{equation}\label{Gs1}
G^{(1)}[\hat{h}^R] = S^{(1)},
\end{equation}
where the $C^0$ source $S^{(1)}$ is given throughout the manifold by 
\begin{equation}\label{S1}
S^{(1)} \equiv - G^{(1)}[\hat{h}^S].
\end{equation}

The logic of the above argument has been that \textit{if} one has a metric perturbation $h$ satisfying equation \eqref{G1} and in a $P$-smooth gauge (equation \eqref{hPBG}), \textit{then} the effective source equation \eqref{Gs1} holds.  In practice, we want to proceed in the reverse direction: we wish to solve equation \eqref{Gs1} and thereby obtain an $h$ satisfying equation \eqref{G1} and in a $P$-smooth gauge.  Retracing the steps of the argument in reverse, it is clear this will hold provided the solution $\hat{h}^R$ of equation \eqref{Gs1} is $C^2$ at $r=0$.\footnote{It is clear that $C^2$ solutions exist by the existence of $P$-smooth gauges, proved by construction in the previous section.}  Obtaining such an $\hat{h}^R$  will depend on the initial and/or boundary conditions chosen, as well on as the choice of gauge.

We first discuss the choice of initial and/or boundary conditions for $h^R$.  We view the specification of a ``physical situation of interest'' as a choice of initial and/or boundary conditions for the metric peturbation $h$.  In principle, one would first determine such conditions in a $P$-smooth gauge and then infer the relevant conditions on $\hat{h}^R = h - \hat{h}^S$.  In practice, determining appropriate initial conditions for $h$ is likely to prove difficult, even without the added requirement of using a $P$-smooth gauge.  Faced with difficulty determining appropriate initial data, the usual strategy is simply to choose inappropriate initial data and evolve in the hopes that at a later time (after ``spurious radiation'' has left the system) the solution will nevertheless resemble the desired physical situation.  We suggest that one employ this strategy at the level of the regular field $\hat{h}^R$, where one could simply choose trivial initial data (or a suitable generalization should trivial initial data conflict with any gauge conditions used).  Effective source calculations made with the scalar wave equation \cite{effective-source} suggest that this strategy will prove effective in the gravitational case as well.

We next discuss the choice of gauge.  Since $\hat{h}^R$ is related to the metric perturbation $h$ by addition of a fixed quantity $\hat{h}^S$, the usual arguments that $h$ and $h + \mathcal{L}_v g$ represent the same physical configuration imply that $\hat{h}^R$ and $\hat{h}^R + \mathcal{L}_v g$ represent the same physical configuration.  In particular, any gauge condition that is ``allowed'' for $h$ will remain ``allowed'' for $\hat{h}^R$.  For example, it is well known that one may impose the Lorenz condition on a smooth perturbation $h$, $\nabla^\mu H_{\mu \nu}=0$ with capitalization denoting trace-reversal, $H_{\mu \nu} = h_{\mu \nu} - (1/2) g_{\mu \nu} h$.  Similarly, one may argue identically that it is always possible to impose the Lorenz condition on the regular field, $\nabla^\mu \hat{H}^R_{\mu \nu}=0$, where capitalization denotes trace-reversal.  In this case equation \eqref{Gs1} becomes
\begin{equation}
E_{\mu \nu}[\hat{h}^R]=S^{(1)},
\end{equation}
where $E_{\mu \nu}$ is the Lorenz-gauge linearized Einstein tensor (a well-studied hyperbolic wave operator on $g$),
\begin{equation}\label{waveop}
E_{\mu \nu}[h]=\nabla^\gamma \nabla_\gamma H_{\mu \nu} - 2 R^{\alpha \ \ \beta}_{\ \mu \nu} H_{\alpha \beta}.
\end{equation}
where the capitalization of the arbitrary perturbation $h$ represents trace-reversal.  Since $E_{\mu \nu}$ is a hyperbolic wave operator, it is expected that the $C^0$ source $S^{(1)}$ will give rise to a $C^2$ solution $\hat{h}^R$,\footnote{While general theorems on wave operators (e.g., prop. 7.4.7 of \cite{hawking-ellis}) would guarantee only weaker regularity of the solution, experience with the effective source method for scalar wave operators \cite{effective-source} shows that sources of our type do in practice give rise to sufficiently regular solutions.} and therefore that the gauge condition $\nabla^\mu \hat{H}^R_{\mu \nu}$ does in fact provide a metric perturbation $h=\hat{h}^R + \hat{h}^S$ in a $P$-smooth gauge.  Note, however, that this gauge differs from ``the Lorenz gauge'', which refers to the Lorenz condition on the full pertubation $h$, $\nabla^\mu H_{\mu \nu}=0$.  If one desires this condition to be satisfied, one instead needs to enforce $\nabla^\mu \hat{H}^R_{\mu \nu} = - \nabla^\mu \hat{H}^S_{\mu \nu}$, in which case equation \eqref{Gs1} becomes
\begin{equation}\label{hlorenz}
E_{\mu \nu}[h^R]=S^{(1)}_{\mu \nu} + \nabla_{(\mu} \nabla^\alpha \hat{H}^S_{\nu) \alpha},
\end{equation}
where capitalization denotes trace-reversal.  We see that the failure of $\hat{h}^S$ to satisfy the Lorenz condition appears as an extra effective source for $\hat{h}^R$.  We have chosen our $h^S$ to satisfy the Lorenz condition, $\nabla^\mu H^S_{\mu \nu} = O(r^2)$, so that the failure comes only from the choice of extension, and the the right-hand-side of \eqref{hlorenz} remains $C^0$.  In particular, the solution $\hat{h}^R$ should be $C^2$, so that the Lorenz gauge is $P$-smooth.  (Indeed, we chose the $P$ gauge and hence the $P$-smooth class precisely so that Lorenz gauges would be included.)  However, we emphasize that while it may be useful to use the Lorenz gauge to compare with previous work or to determine a first-order-accurate long-term evolution via a particular proposed prescription (see discussion in section \ref{sec:inspiral}), for the purposes of determining $h$ there is no fundamental reason to prefer one gauge over another.

In implementing the effective source method, above, we have made a convenient choice of ``singular field'' $h^S$.  However, we emphasize that many other choices could have been made, with equivalent results.  In particular, one may modify $h^S$ by the addition of any given smooth function $f$.  In this case the effective source $-G^{(1)}[\hat{h}^S]$ will remain $C^0$ (though it will no longer be $O(r)$, since $h^S$ is no longer a solution to all orders considered), and the full metric pertubation $h=\hat{h}^R+\hat{h}^S$ will remain the same, provided the appropriate initial/boundary/gauge conditions for the corresponding new $\hat{h}^R$ (modified by $-f$) are chosen.  However, while we delay a systematic discussion of the motion until section \ref{sec:motion}, we note here that an advantage of our particular choice of $h^S$ is that the first-order motion may be described as geodesic in the peturbation $\hat{h}^R$ (as in the original treatment of \cite{detweiler-whiting}).  The basic point is that, as may be seen from equation \eqref{hRext2} with $\xi=0$ and equation \eqref{hH}, we have $\hat{h}^R=O(r^2)$ in the $P$ gauge, i.e., the regular field and its first derivative vanish on the worldline.  The statement that the perturbed motion vanishes (together with the statement that the background motion is geodesic) may then be expressed equivalently as the statement of geodesic motion in $g+\lambda \hat{h}^R$, which, as a covariant statement, will hold in any smoothly related gauge.  This argument is given more formally and explicitly in section \ref{sec:motion}, below. 

\subsection{Determination of $\delta \mathcal{E}_{ij}$, $\delta \mathcal{B}_{ij}$, and $\xi_\mu$.}\label{sec:PBG}

Our next task is the identification of an appropriate singular field at second order.  At first order, the singular field was found by noting that the unknown tensors $\{\delta \mathcal{E}_{ij}, \delta \mathcal{B}_{ij}\}$ appeared in the $P$-gauge perturbation only in a smooth way, so that a singular part depending only on the known tensors $\{\mathcal{E}_{ij},\mathcal{B}_{ij},\mathcal{E}_{ijk},\mathcal{B}_{ijk}\}$ could be chosen.  Furthermore, since smooth gauge transformations affect the metric perturbation only by addition of a smooth term, this choice of singular field guarantees that $h-h^S$ is regular in all $P$-smooth gauges.  At second order, however, the unknown tensors $\{\delta \mathcal{E}_{ij}, \delta \mathcal{B}_{ij}\}$ do appear  as part of singular terms (see equation \eqref{jP}, where the $r a_{32}$ terms are not differentiable).  Furthermore, an identification of a singular part, $j^S$, of $j^P$ does not guarantee that $j-j^S$ is regular in all $P$-smooth gauges, since smooth gauge transformations change the second-order metric perturbation by a singular term, $\mathcal{L}_\xi h^P$ (see equation \eqref{g2t} and recall that $h^P$ is singular).  To correctly identify a singular part of $j$ will therefore require expressions for all of the unknown quantities $\{\delta \mathcal{E}_{ij}, \delta \mathcal{B}_{ij}, \xi_\mu\}$ that appear in the expression for the general $P$-smooth metric perturbation, $h=h^P-\mathcal{L}_\xi g$.  

The relevant question is the following: given a perturbation $h$ in a $P$-smooth gauge (imagined, e.g., to have been numerically computed by the prescription given in the previous section), how can we express this perturbation as $h=h^P-\mathcal{L}_\xi g$ for some $\{\delta \mathcal{E}_{ij}, \delta \mathcal{B}_{ij}, \xi_\mu\}$?  (We remind the reader that $h^P$ is constructed from $\delta \mathcal{E}_{ij}$ and $\delta \mathcal{B}_{ij}$.)  Below we find  that there is precisely a ten-parameter freedom in the choice of $\{\delta \mathcal{E}_{ij}, \delta \mathcal{B}_{ij}, \xi_\mu\}$ that specifies a decomposition of the form $h=h^P-\mathcal{L}_\xi g$, and give a prescription for computing these quantities in terms of an integration of transport equations along $\gamma$.  As shown therein, knowledge of $\xi$ determines $\delta \mathcal{E}_{ij}$ and $\delta \mathcal{B}_{ij}$, so that we may view the ten-parameter freedom in the decomposition as a ten-parameter freedom in the choice of $\xi$.  Since different such choices lead to different second-order metric perturbations but (by construction) preserve the first-order perturbation $h$, this freedom corresponds to the influence of first-order gauge freedom on the second-order metric perturbation.  The freedom in choice of $\delta \mathcal{E}_{ij}$,  $\delta \mathcal{B}_{ij}$ and $\xi$ (at fixed $h$) may be viewed as first-order gauge freedom that manifests only at second order in the metric components.

Since $h^S$ is a specified function of known quantities, we may without loss of generality consider the smooth vacuum perturbation $h^R=h-h^S$.  (Since we will work with the regular field only locally near $r=0$ in this section, we drop the hat in its notation.) From equation \eqref{hRext} we have
\begin{equation}\label{hR}
h^R = h^H - \mathcal{L}_\xi g + O(r^3),
\end{equation}
where we remind the reader that $h^H$ is a simple function of $\delta \mathcal{E}_{ij}$ and $\delta \mathcal{B}_{ij}$, given by equation \eqref{hH}.  To show that $\delta \mathcal{E}_{ij}$ and $\delta \mathcal{B}_{ij}$ (and hence $h^H$) are determined by $\xi$, we compute the linearized Riemann tensor of $h^R$, yielding 
\begin{align}
R^{(1)}_{\alpha \beta \gamma \delta}[h^R] & = R^{(1)}_{\alpha \beta \gamma \delta}[h^H]  + R^{(1)}_{\alpha \beta \gamma \delta}[- \mathcal{L}_\xi g] + O(r) \nonumber \\
& = R^{(1)}_{\alpha \beta \gamma \delta}[h^H] - \mathcal L_\xi R_{\alpha \beta \gamma \delta} + O(r).\label{R1xi}
\end{align}
(The definition of $R^{(1)}_{\mu \nu \rho \sigma}[h]$ is given below equation \eqref{deltaB}.) In the second line we have used the covariance property $F^{(1)}
[\mathcal{L}_\xi g]=\mathcal{L}_\xi F^{(0)}$, holding for any covariant function of the metric $F[g]$.  Using equations \eqref{deltaE} and \eqref{deltaB} we then have
\begin{align}
\delta \mathcal{E}_{ij} & = \left. \left( R^{(1)}_{0 i 0 j}[h^R] + \mathcal L_\xi R_{0 i 0 j} \right)\right|_{\gamma} \label{deltaExi} \\
\delta \mathcal{B}_{ij} & = -\frac{1}{2} \epsilon^{kl}_{\ \ i} \left. \left( R^{(1)}_{0 j k l}[h^R] + \mathcal L_\xi R_{0 j k l} \right) \right|_{\gamma}, \label{deltaBxi}
\end{align}
where our Lie derivative expressions refer to components of Lie derivatives of the (background) Riemann tensor (rather than some kind of derivative of a component).  Thus $\delta \mathcal{E}_{ij}$ and $\delta \mathcal{B}_{ij}$ (and hence $h^H$) are determined by $\xi$ and its first derivative on the worldline.  In particular, introducing tensors $A_{\mu}$ and $B_{\mu \nu}$ defined along $\gamma$ by
\begin{align}
A_{\mu} & = \xi_\mu|_\gamma \label{Adef} \\
B_{\mu \nu} & = (\nabla_\nu \xi_\mu)|_\gamma, \label{Bdef}
\end{align}
we have
\begin{align}
\delta \mathcal{E}_{ij} & = \frac{1}{2} \left( -\partial_i \partial_j h^R_{00} + 2 \partial_0 \partial_{(i} h^R_{j)0} - \partial_0 \partial_0 h^R_{ij} \right) +h^R_{00} \mathcal{E}_{ij}\nonumber \\
& + 2 B_{k0} \epsilon^k_{\ l(j} \mathcal{B}_{i)}^{\ \ l} + \frac{2}{3} A_k \epsilon^k_{\ l(j} \dot{\mathcal{B}}_{i)}^{\ \ l} + 2B_{k(i}\mathcal{E}_{j)}^{\ k} \nonumber \\ & - A_0 \dot{\mathcal{E}}_{ij} + A_k \mathcal{E}_{ij}^{\ \ k} \label{deltaExicoord} \\
\delta \mathcal{B}_{ij} & = \frac{1}{2} \epsilon_{i}^{\ kl} \left( \partial_j \partial_l h^R_{0k} - \partial_k \partial_l h^R_{0j} \right) + \delta^{kl}B_{kl} \mathcal{B}_{ij} + \frac{1}{2}h^R_{00} \mathcal{B}_{ij} \nonumber \\ & + 2 B_{k[j} \mathcal{B}_{i]}^{\ \ k} -A_0 \dot{\mathcal{B}}_{ij} + \frac{8}{3} A_k \mathcal{B}_{ij}^{\ \ k} - 2 \epsilon^k_{\ li} \mathcal{E}_j^{\ l} B_{k0} \nonumber \\ & - \epsilon^k_{\ l i} h^R_{0k} \mathcal{E}_j^{\ l} + B_{k0}\epsilon_{ijl} \mathcal{E}^{kl} - \frac{2}{3} A_k \epsilon^k_{\ li} \dot{\mathcal{E}}_j^{\ l} + \frac{1}{3} A_k \epsilon_{ijl} \dot{\mathcal{E}}^{kl} \label{deltaBxicoord} 
\end{align}
(Note that while $\delta\mathcal{B}_{ij}$ is not given above in manifestly symmetric form, one may easily confirm its symmetry using the fact that $h^R$ is a vacuum perturbation.)  

We now regard equation \eqref{hR} as an equation for $\xi$,
\begin{equation}\label{xieq}
-2 \nabla_{(\mu} \xi_{\nu)} = h^R_{\mu \nu} - h^H_{\mu \nu} + O(r^3),
\end{equation}
where $h^H$ is constructed from $\xi$ via equations \eqref{deltaExi}-\eqref{deltaBxi} (equivalently \eqref{Adef}-\eqref{deltaBxicoord}) and \eqref{hH}.  Taking a derivative and employing manipulations normally used for Killing's equation (e.g., appendix C of \cite{wald}), we have
\begin{equation}\label{gradgradxi}
\nabla_\alpha \nabla_\beta \xi_\gamma + R_{\beta \gamma \alpha}^{\ \ \ \ \delta} \xi_{\delta} = - \Gamma^{(1)}_{\gamma \alpha \beta}[h^R-h^H] + O(r^2),
\end{equation}
where $\Gamma^{(1)}[h]$ is the perturbed Christoffel symbol with a lowered index,
\begin{equation}
\Gamma^{(1)}_{\gamma \alpha \beta}[h] = \frac{1}{2} \left( \nabla_\alpha h_{\beta \gamma} + \nabla_\beta h_{\alpha \gamma} - \nabla_\gamma h_{\alpha \beta}  \right).
\end{equation}
Equation \eqref{gradgradxi} shows that solutions to equation \eqref{xieq} are determined everywhere by a choice of $\xi$ and $\nabla \xi$ at a single point.  Equation \eqref{xieq} restricts this choice to a ten-parameter-family (such as ``Killing data'' $\xi_\mu$ and $\nabla_{[\mu}\xi_{\nu]}$).  We now show constructively that all such choices lead to solutions to equation \eqref{xieq}.

Since $\Gamma^{(1)}[h^H]$ is $O(r)$, equations \eqref{xieq} and \eqref{gradgradxi} give for $A_{\mu}$ and $B_{\mu \nu}$ that
\begin{align}
B_{(\mu \nu)} &= -\frac{1}{2} h_{\mu \nu}^R|_\gamma \label{fixB} \\
u^\alpha \nabla_\alpha A_\mu & = B_{\mu \alpha} u^\alpha \label{transA} \\
u^\alpha \nabla_\alpha B_{\mu \nu} & = R_{\mu \nu \alpha \beta} u^\alpha A^\beta -  u^\alpha \Gamma^{(1)}_{\mu \alpha \nu}[h^R]|_\gamma. \label{transB}
\end{align}
Equations \eqref{transA}-\eqref{transB} give transport rules for $A$ and $B$ along $\gamma$, while equation \eqref{fixB} gives a constraint (which is preserved by the transport.)  In the RWZ coordinates these may be written
\begin{align}
\dot{A}_0 & = -\frac{1}{2} h^R_{00} \label{transAcoord1} \\
\dot{A}_i & = B_{i0} \label{transAcoord2} \\
\dot{B}_{i0} & = - \mathcal{E}_{ij} A^j - \partial_0 h^R_{0i} + \frac{1}{2}\partial_i h^R_{00} \label{transBcoord1} \\
\dot{B}_{[ij]} & = \epsilon_{ijl} \mathcal{B}^l_{\ k} A^k + \partial_{[i}h^R_{j]0}, \label{transBcoord2}
\end{align}
where the overdot denotes a $t$ derivative and $h^R_{\mu \nu}$ and its derivatives are evaluated at $x^i=0$ (i.e., on  $\gamma$).  Equations \eqref{transAcoord1}-\eqref{transBcoord2} (together with \eqref{fixB}) determine $A_\mu$ and $B_{\mu \nu}$ given a choice of initial data for $\{A_\mu,B_{i0},B_{[ij]}\}$.  The reader may recognize the last two terms of equation \eqref{transBcoord1} as the self-force on the particle, here taking a ``perturbed geodesic equation'' form.  As discussed in more detail in our systematic treatment of the motion in section \ref{sec:motion} below, our definition of motion implies that $Z^{(1)\mu}=\xi^\mu|_\gamma=A^\mu$, so that the transport equation for $A^\mu$ is in fact the first-order equation of motion.  However, in the present section we confine ourselves to the derivation of a prescription for computing the first and second order metric perturbation, for which the interpretation of $\xi^\mu|_\gamma$ as giving the motion is entirely irrelevant.

Given a choice of $\{A_\mu,B_{i0},B_{[ij]}\}$ at some point along $\gamma$, equations \eqref{transAcoord1}-\eqref{transBcoord2} determine these quantities everywhere on $\gamma$.  We now imagine that a choice has been made, so that $A$ and $B$ are known along $\gamma$.  This determines $\delta \mathcal{E}_{ij}$ and $\delta \mathcal{B}_{ij}$ via equations \eqref{deltaExicoord}-\eqref{deltaBxicoord} (equivalently \eqref{deltaExi}-\eqref{Bdef}) and hence $h^H$ by equation \eqref{hH}.   The right hand side of equation \eqref{gradgradxi} is then ``known'' in terms of the value and derivative of $\xi$ on the worldline (i.e., in terms of $A$ and $B$), so that we may determine $\xi$ to higher order in $r$ by expanding the left hand side in $r$ and equating orders in $r$.  After some effort we obtain
\begin{widetext}
\begin{align}
%xi0
\xi_0 & = A_0 - B_{i0}x^i - h^R_{0i} x^i + \left( - \frac{1}{2} \partial_j h^R_{0i} + \frac{1}{4} \partial_0 h^R_{ij} + A_0 \mathcal{E}_{ij} + A^k \epsilon_{kil} \mathcal{B}_{j}{}^{l} \right) x^i x^j + \left( - \frac{1}{6} B_{j0} \mathcal{E}_i^{\ j} + \frac{5}{18} A_j  \dot{\mathcal{E}}_i^{\ j} \right) x^i r^2 \nonumber \\ 
%xi0: x x x 
& + \left( - \frac{1}{6} \partial_j \partial_k h^R_{0i} + \frac{1}{12}
\partial_0 \partial_k h^R_{ij}  + \frac{2}{3} B_{lj} \epsilon^l_{\ im} \mathcal{B}_{k}{}^{m} + \frac{8}{9} A^l \epsilon_{lj}{}^{m} \mathcal{B}_{kim} 
- \frac{1}{3} B_{i0} \mathcal{E}_{jk} - \frac{2}{3} h^R_{0i} \mathcal{E}_{jk}  - \frac{4}{9} A_i \dot{\mathcal{E}}_{jk} + \frac{1}{3} A_0 \mathcal{E}_{ijk} \right) x^i x^j x^k \nonumber \\
& + O(r^4)  \label{xi0} \\ 
%xii
\xi_i & = A_i + B_{ij} x^j - \left( \frac{1}{2} \partial_k h^R_{ij} + \frac{1}{4} \partial_i h^R_{jk} - A_i \mathcal{E}_{jk} - \frac{2}{3} A_0 \epsilon_{ikl} \mathcal{B}_{j}{}^{l} \right) x^j x^k + \left( n_i A_j n_{k} \mathcal{E}^{jk} - \frac{1}{2} 
A_j \mathcal{E}_i^{\ j} \right) r^2 \nonumber \\
%xii: x r^2
 & + \left( \frac{1}{12} \partial_j \partial_0 h^R_{0i} + \frac{1}{12} \partial_i \partial_0 h^R_{0j}  - \frac{1}{12} \partial_i \partial_j h^R_{00} - \frac{1}{12} \partial_0 \partial_0 h^R_{ij} + \frac{1}{6} h^R_{00} \mathcal{E}_{ij} + \frac{1}{6} h^R_{ik} \mathcal{E}_{j}^{\ k} - \frac{1}{6} h^R_{ik} \mathcal{E}_{j}^{\ k}\right) r^2 x^j \nonumber \\ 
& + \left( - \frac{1}{6} B_{k0} \epsilon^k_{\ jl} \mathcal{B}_{i}{}^{l} + \frac{1}{6} B_{k0} \epsilon_{i \ l}^{\ k} \mathcal{B}_{j}{}^{l} - \frac{1}{12} A_k \epsilon^k_{\ jl} \dot{\mathcal{B}}_{i}{}^{l} - \frac{1}{12} A_{k} \epsilon_{ijl} \dot{\mathcal{B}}^{kl} \right) r^2 x^j \nonumber \\ 
%xii: x x r
& + \left( - \frac{1}{3} \partial_0 \partial_k h^R_{0j} + \frac{1}{6} \partial_j \partial_k h^R_{00} + \frac{1}{6} \partial_0 \partial_0 h^R_{jk} + \frac{2}{3} A_0 \dot{\mathcal{E}}_{jk} - \frac{1}{3} h^R_{00} \mathcal{E}_{jk} + \frac{2}{3} B_{l0} \epsilon^l_{\ jm} \mathcal{B}_{k}{}^{m} + \frac{5}{12} A_l \epsilon^l_{\ km} \dot{\mathcal{B}}_{j}{}^{m} \right) x_i x^j x^k \nonumber \\
%xii: x x x
 & + \left( - \frac{1}{6} \partial_k \partial_l h^R_{ij} + \frac{1}{12} \partial_i \partial_l h^R_{jk} - \frac{1}{3} h^R_{ij} \mathcal{E}_{kl} + \frac{1}{3} \epsilon_{ilm} h^R_{0j} \mathcal{B}_{k}{}^{m} \right) x^j x^k x^l \nonumber \\ & + \left( \frac{1}{3} B_{j0} \epsilon_{ilm} \mathcal{B}_{k}{}^{m} + \frac{1}{12} A_j \epsilon_{ilm} \dot{\mathcal{B}}_{k}{}^{m}  - \frac{2}{3} A_0 \epsilon_{ij}{}^{m} \mathcal{B}_{klm} - \frac{1}{3} B_{jk} \mathcal{E}_{il}  - B_{ij} \mathcal{E}_{kl} - \frac{1}{3} A_i \mathcal{E}_{jkl} \right) x^j x^k x^l + O(r^4) \label{xii},
\end{align}
\end{widetext}
where $h^R$ and its derivatives are evaluated on $\gamma$.  We then check by direct computation that the above formula does give a solution to equation \eqref{xieq} (and not just \eqref{gradgradxi}), provided that $A$ and $B$ satisfy the transport equations \eqref{transAcoord1}-\eqref{transBcoord2} and that $h^R$ is a vacuum perturbation.  Thus equations \eqref{transAcoord1}-\eqref{xii}  provide a ten-parameter-family of solutions for $\xi$ to equation \eqref{xieq} and hence \eqref{hPBG}.  Since it was already shown that the general solution is at most a ten-parameter-family, the general solution is in fact a ten-parameter family, and all solutions may be constructed this way.

The main results of this subsection are equations \eqref{deltaExicoord}, \eqref{deltaBxicoord}, and \eqref{xi0}-\eqref{xii}, which give expressions for $\delta \mathcal{E}_{ij}$, $\delta \mathcal{B}_{ij}$ and $\xi_\mu$ in terms of an integration of the transport equations \eqref{transAcoord1}-\eqref{transBcoord2} for $A_\mu$ and $B_{\mu \nu}$.  We have used the tensor analysis package \textit{xTensor} \cite{xTensor} for the software package \textit{Mathematica} \cite{mathematica} to perform the extensive computations of this subsection.

\subsection{Second Order Effective Source}\label{sec:secondorder}

Equation \eqref{jP} gives the general second-order metric perturbation in series in $r$ in a particular gauge (the ``P gauge'').  In a smoothly related gauge, the second-order metric perturbation is given by (see equations \eqref{gauge-coord} and \eqref{g2t})
\begin{equation}\label{jPBG}
j = j^P - \mathcal{L}_{\xi} h^P + \frac{1}{2} \left( \mathcal{L}_\xi \mathcal{L}_\xi g - \mathcal{L}_\Xi g \right).
\end{equation}
Since a prescription for computing $\delta \mathcal{E}_{ij}$, $\delta \mathcal{B}_{ij}$, and $\xi$ has now been given, the first two terms on the right-hand-side may be considered ``known''.  Since the remaining terms are regular, an appropriate singular field is thus
\begin{equation}\label{jS}
j^S = j^P - \mathcal{L}_\xi h^P.
\end{equation} 
One may now straightforwardly combine equations \eqref{jP}, \eqref{hP}, \eqref{deltaExicoord}-\eqref{deltaBxicoord}, and \eqref{xi0}-\eqref{xii} to produce an expression for $j^S$ in terms of $\mathcal{E}_{ij}$, $\mathcal{B}_{ij}$, $\mathcal{E}_{ijk}$, $\mathcal{B}_{ijk}$, $A_\mu$, and $B_{\mu \nu}$.  This expression is given in equations \eqref{jS00}-\eqref{jSij} of appendix \ref{sec:appsing}.  We remind the reader that the choice of initial data for $A_\mu$ and $B_{\mu \nu}$ constitutes a choice of first-order gauge freedom that manifests only at second (and higher) order.  In particular,  $A_\mu$ represents the perturbed position of the particle, and in this sense the second-order singular field---and hence effective source---``knows'' about the first-order deviation from geodesic motion.

Following the same logic as in the first-order case, one should compute $j^S$ to $O(r)$ and then choose an arbitrary extension, $\hat{j}^S$, to the entire manifold (minus $\gamma$).  We then introduce a regular field $\hat{j}^R$ by
\begin{equation}\label{jRext}
\hat{j}^{R} = j-\hat{j}^S,
\end{equation}
and plug in to the second-order Einstein equation \eqref{G2} to get
\begin{equation}\label{G2sint}
G^{(1)}[\hat{j}^R] = -G^{(1)}[\hat{j}^S] - G^{(2)}[h] \qquad (\textrm{for } r>0).
\end{equation}
While each term on the right-hand-side of equation \eqref{G2sint} blows up at $r=0$, by construction the sum is $O(1)$.\footnote{To see this explicitly, note that equation \eqref{G2sint} holds to $O(r^{-1})$ if the hats are removed: $G^{(1)}[j^R + j^S + O(r^2)]=-G^{(2)}[h]$.  It then follows that  $G^{(1)}[j^S]+G^{(2)}[h]=-G^{(1)}[j^R]+O(1)=O(1)$.} Thus the right-hand-side is in fact bounded (but not necessarily continuous) at $r=0$.  We may nevertheless drop the requirement that $r>0$ by interpreting \eqref{G2sint} in a Sobalev (or distributional\footnote{To give a distributional interpretation we promote the entire right-hand-side of \eqref{G2sint} to a distribution.  We give no distributional interpretation to each term separately.}) sense.  We therefore write 
\begin{equation}\label{G2s}
G^{(1)}[\hat{j}^R] = S^{(2)}
\end{equation}
with
\begin{align}\label{S2}
S^{(2)} & \equiv - G^{(1)}[\hat{j}^S] - G^{(2)}[h]
\end{align}
where the effective source is bounded but potentially discontinuous.  

As at first order, one may determine the perturbation $j$ in a P-smooth gauge by solving equation \eqref{G2s} with initial, boundary, and/or gauge conditions such that $\hat{j}^R$ is sufficiently regular (in this case $C^1$), and it appears that the Lorenz condition on the regular field, $\nabla^\mu \hat{J}^R_{\mu \nu}=0$, (where capitalization denotes trace-reversal) should be an appropriate gauge choice.\footnote{Unlike at first order, however, it is not possible to impose the Lorenz condition on the full metric perturbation $j$ by our effective source method, since our second-order singular field violates the Lorenz condition by a singular amount.  (The analog of equation \eqref{hlorenz} would then contain a singular source term.)  More discussion of this gauge condition can be found in the text below equation \eqref{a32ij}.}

\section{Motion in P-smooth gauges}\label{sec:motion}

We have now developed a prescription for computing the global metric perturbation in $P$-smooth gauges, where (by definition) the metric may be written
\begin{align}
h & = h^P - \mathcal{L}_\xi g \label{hgen} \\
j & = j^P - \mathcal{L}_\xi h^P + \frac{1}{2} \left( \mathcal{L}_\xi \mathcal{L}_\xi g -  \mathcal{L}_\Xi g \right), \label{jgen}
\end{align}
for smooth $\xi$ and $\Xi$.  Since the motion is defined to vanish in $P$ gauge, the motion in the above $P$-smooth gauge is given by (see equation \eqref{Z2t})
\begin{align}
Z^{(1)\mu} & = \xi^\mu|_\gamma \label{Z1} \\
Z^{(2)\mu} & = (\Xi^\mu+\xi^\nu \partial_\nu \xi^\mu)|_\gamma. \label{Z2}
\end{align}
Recall that we previously notated $\xi_\mu|_\gamma$ by $A_\mu$ (see equation \eqref{Adef}).  Thus our $A_\mu$ in fact gives the first-order motion, and the analysis of section \ref{sec:PBG} has in fact produced the first-order equation of motion in equations \eqref{transAcoord1}-\eqref{transBcoord1}.  From the point of view of the systematic calculations performed there, it comes as some surprise that the form of perturbed geodesic equation emerges.  We now use a simple (and trivial) argument to show why the form of the perturbed geodesic equation must in fact occur.  This argument also derives the second-order equation of motion in terms of a second-perturbed geodesic form.

The argument proceeds as follows.  In the P gauge, the description of motion is geodesic in the background metric $g$ (since the perturbed motions vanish).  To determine the description in smoothly related gauges, use the P gauge to promote the background metric to a finite-$\lambda$ tensor, $g^{BG}_{\mu \nu}(\lambda)\equiv g^{(0)}_{\mu \nu}$, where this equation holds only in the P gauge.  Within the class of P-smooth gauges, one now has the invariant description of motion that $Z^{\mu}(\lambda)$ is geodesic in $g_{\mu \nu}^{BG}(\lambda)+O(\lambda^3)$.  Perturbatively, we have
\begin{equation}\label{gBG}
g^{BG}(\lambda) = g - \lambda h^{BG} + \lambda^2 j^{BG} + O(\lambda^3),
\end{equation}
with
\begin{align}
h^{BG} & = -\mathcal{L}_{\xi} g \label{hBG} \\ 
j^{BG} & = \frac{1}{2}(\mathcal{L}_\xi \mathcal{L}_\xi  g - \mathcal{L}_{\Xi} g), \label{jBG}
\end{align}
and it follows that the first and second perturbed positions $Z^{(1)}$ and $Z^{(2)}$ must satisfy the first and second perturbed geodesic equation in first and second perturbations $h^{BG}$ and $j^{BG}$.  At first order, we have already found that the motion is given by the perturbed geodesic equation in our regular field $h^R$.  But from (e.g.) equation \eqref{hR} we have
\begin{align}
h^R & = h^H + h^{BG} + O(r^3) \label{hRHBG} \\
 & = h^{BG} + O(r^2), \label{hRBG}
\end{align}
where the second line follows from the fact that $h^H=O(r^2)$.  Since the perturbed geodesic equation (equation \eqref{Z1g}) includes only first spatial derivatives of the perturbation, equation \eqref{hRBG} shows that the statement of geodesic motion in $h^R$ is equivalent to the statement of geodesic motion in $h^{BG}$.  This ``explains'' the appearance of the geodesic form in equation \eqref{transA}, and suggests that the motion is more naturally regarded as geodesic in $h^{BG}$ (which happens to coincide with our choice of $h^R$ to the relevant order).  This viewpoint has fundamental appeal in that the motion, which is pure gauge, is given in terms of a pure gauge metric perturbation.

To determine the second-order equation of motion, we could similarly proceed to directly ``solve'' equation \eqref{jgen} for $\Xi$, as we did in section \ref{sec:PBG} to solve equation \eqref{hgen} for $\xi$ (though our goal there was the formulation of a second-order effective source).  However, we may avoid this task by appealing to the above argument, which shows that the second-perturbed description of motion is the second-perturbed geodesic equation in perturbations $h^{BG}$ and $j^{BG}$.  From equations \eqref{jgen}, \eqref{jS} and \eqref{jRext}, we have for our particular choice of $j^R$ that
\begin{equation}\label{jRBG}
j^R = j^{BG} + O(r^2).
\end{equation}
Since the second-perturbed geodesic equation (equation \eqref{Z2g}) contains only first spatial derivatives of the second perturbation, we may equally well use $j^R$ instead of $j^{BG}$ in determining the motion.  However, the second-perturbed geodesic equation also contains a term involving the \textit{second} spatial derivative of the first perturbation, and for this term the difference between $h^R$ and $h^{BG}$ is relevant, since these quantities agree only to $O(r)$ (see equation \eqref{hRHBG}).  To solve for the second-order motion one must first determine $h^{BG}$.  This may be accomplished by subtracting $h^H$ from $h^R$ (see equation \eqref{hRHBG}), where $h^H$ may be determined from equation \eqref{hH} with \eqref{deltaExicoord}-\eqref{deltaBxicoord}.  The motion is then given by solving equation \eqref{Z2g} with $g^{(1)} \rightarrow h^{BG}$ and $g^{(2)}\rightarrow j^{BG}$.

Note, however, that the term relevant for the difference between $h^R$ and $h^{BG}$ is simply  $(1/2) Z^{(1)j} \partial_j \partial_i h^{BG}_{00}$ (appearing in the second line of the expression for $\ddot{Z}^{(2)}_{\ \ \ i}$ in equation \eqref{Z2g}), and so it is in fact only necessary to consider the $00$ component of $h^{BG}$.  In particular, we have
\begin{equation}\label{large-panda}
h^{BG}_{00} = h^{R}_{00} - h^H_{00} = h^R_{00} + \delta \mathcal{E}_{ij} x^i x^j + O(r^3),
\end{equation}
where $\delta \mathcal{E}_{ij}$ is given by equation \eqref{deltaExicoord}.

Although we view the interpretation of geodesic motion in the $BG$ fields as having more fundamental status, it is the regular fields that will arise in practice, and we now explicitly present the final equations of motion in terms of the regular fields.  Using equations \eqref{hRHBG}, \eqref{jRBG} and \eqref{large-panda} to relate $\{h^{BG},j^{BG}\}$ to $\{h^R,j^R\}$, the final equations of motion (equations \eqref{Z1g}-\eqref{Z2g} with $g^{(1)} \rightarrow h^{BG}$ and $g^{(2)}\rightarrow j^{BG}$) become
\begin{align}
\ddot{Z}^{(1)}_{\ \ \ 0} & = - \frac{1}{2} \partial_0 h^R_{00} \nonumber \\
\ddot{Z}^{(1)}_{\ \ \ i} & = - \partial_0 h^R_{0i} + \frac{1}{2} \partial_i h^R_{00} - \mathcal{E}_{ij} Z^{(1)j} \label{Z1hR}
\end{align}
and
\begin{align}
\ddot{Z}^{(2)}_{\ \ \ 0} & = - \frac{1}{2} \partial_0 j^R_{00} - h^R_{0\nu} \ddot{Z}^{(1)\nu} \nonumber \\ & + \dot{Z}^{(1)\gamma} \partial_{\gamma} h^R_{00} + \frac{1}{2} Z^{(1)\gamma} \partial_\gamma \partial_0 h^R_{00} \nonumber \\ & - 2 \mathcal{E}_{ij}\dot{Z}^{(1)i}Z^{(1)j} - \frac{1}{2} \dot{\mathcal{E}}_{ij} Z^{(1)i} Z^{(1)j}, \nonumber \\
\ddot{Z}^{(2)}_{\ \ \ i} & = - \partial_0 j^R_{0i} + \frac{1}{2} \partial_i j^R_{00} - \mathcal{E}_{ij} Z^{(2)j} + \delta \mathcal{E}_{ij} Z^{(1)j} \nonumber \\ & - h^R_{i \nu} \ddot{Z}^{(1)\nu} + Z^{(1)\gamma} \partial_{\gamma} (-\partial_0 h^R_{0i} + \frac{1}{2} \partial_i h^R_{00}) \nonumber \\ & + 2 \dot{Z}^{(1)0} \ddot{Z}^{(1)i} - \dot{Z}^{(1)j} \left( \partial_0 h^R_{ij} + \partial_j h^R_{i0} - \partial_i h^R_{j0}\right) \nonumber \\ & -2\dot{Z}^{(1)j}Z^{(1)k} \epsilon_{ijl}\mathcal{B}_k^{\ l} + \frac{2}{3} Z^{(1)k} Z^{(1)l} \epsilon_{ipk} \dot{\mathcal{B}}_l^{\ p} \nonumber \\ & - \frac{1}{2} \mathcal{E}_{ijk} Z^{(1)j} Z^{(1)k} - \dot{\mathcal{E}}_{ij} Z^{(1)0}Z^{(1)j} \label{Z2hR},
\end{align}
where $\delta \mathcal{E}_{ij}$ is given by equation \eqref{deltaExicoord}.  In this form, the second-order equation of motion is seen to be geodesic in the regular fields, up to a correction term (the term proportional to $\delta \mathcal{E}_{ij}$) that accounts for the fact that the motion is in fact only geodesic in the $BG$ fields.\footnote{Note that $\delta \mathcal{E}_{ij}$ does not represent the perturbed Riemann tensor of $h^R$ but rather that of $h^H$, which is related to that of $h^R$ by equations \eqref{R1xi}-\eqref{deltaBxi} (see also \eqref{large-panda}).  This accounts for the positive sign in front of the $\delta \mathcal{E}_{ij}$ term.}

\section{Choice of $P$-gauge}\label{sec:choiceP}

The content of this paper has been the identification of a class of gauges for which the motion may be sensibly defined and the development of a prescription for computing the metric and motion in such gauges.  This class was chosen by constructing a particular mass-centered gauge (called $P$ gauge) and considering the class of all gauges related by smooth first and second order gauge vectors ($P$-smooth gauges).  In constructing the $P$ gauge many particular choices were made, and the reader may wonder the effect of making different choices, leading to a $P'$ gauge and possibly distinct class of $P'$-smooth gauges.

Suppose that the content of section \ref{sec:Pgauge} were repeated, except that a different mass-centered gauge, called $P'$ gauge, were chosen.  For concreteness, the reader may imagine that we chose Cartesian Schwarzschild coordinates rather than Cartesian isotropic coordinates for the near-zone background metric, and did not make the additional first-order gauge transformation, equations \eqref{v0}-\eqref{vi}.  This would produce a $P'$ gauge that is related to our $P$ gauge by a first-order far-zone gauge vector of the form $V^i=n^i+O(r)$ (as well as by analogous second and third order gauge vectors), which modifies the metric perturbation by a singular amount (changing the structure of $\mathcal{M}^{(1)}$ from isotropic-type to Schwarzschild-type).   After identifying an appropriate singular field (one option would be transforming the old singular field by $V^\mu$), one could develop an effective source method to determine the metric perturbations in $P'$-smooth gauges. Since the $P'$ gauge is mass-centered, the analysis of the motion will then proceed identically, leading to a prescription for determining the motion, $\{Z'^{(1)},Z'^{(2)}\}$, in $P'$-smooth gauges.

It is clear that the perturbations $\{h,j\}$ and $\{h',j'\}$ in $P$-smooth and (respectively) $P'$-smooth gauges thus constructed will differ by a (possibly singular) gauge transformation (provided that the initial data differ by a gauge transformation), and thus represent the ``same physics''.  The reader may further wonder whether $\{Z^{(1)},Z^{(2)}\}$ and $\{Z'^{(1)},Z'^{(2)}\}$ thus constructed also represent the ``same physics''.  However, since the gauge transformation law for a curve, equations \eqref{Z1t} and \eqref{Z2t}, does not make sense in the presence of singular gauge vectors, we have no a priori criterion with respect to which to check this type of gauge covariance property.  Instead, we may view our definition of motion as (in principle) \textit{providing} a generalized gauge transformation law for the motion that ensures that $\{h,Z^{(1)},j,Z^{(2)}\}$ and $\{h',Z'^{(1)},j',Z'^{(2)}\}$ represent the ``same physics''.  For smooth gauge vectors, the law trivially agrees with equations \eqref{Z1t} and \eqref{Z2t}.  For (non-smooth) gauge vectors that link a $P$-smooth gauge to a $P'$-smooth gauge (for particular known choices of $P$ and $P'$), it should be possible to derive such a law by writing the gauge transformation as the composition of a smooth transformation with the singular (but mass-centered-preserving) transformation that relates $P$ gauge to $P'$ gauge; the law is then simply be equations \eqref{Z1t} and \eqref{Z2t} using the smooth transformation.  For general gauges, the situation is clouded by the fact that the allowed form of the gauge transformation is conjectured but not known \cite{gralla-wald-gauge}, and further that (even restricting to the conjectured class) the class of gauges smoothly connected to a mass-centered gauge (i.e., those for which we can define the motion) is not known at second order.  In the appendix of paper I (see also \cite{gralla-wald-gauge}) we obtained some results at first order;\footnote{More precisely, we showed that for first-order gauge vectors of the form $\xi^\mu = F^\mu(t,\vec{n})+O(r)$ for smooth $F^\mu$, the first-order motion changes by $\delta Z^{(1)i}=(3/4\pi) \langle n^j F_j n^i\rangle$, where the angle brackets denote an average over the sphere.} we have not obtained analogous results at second order, where the situation is far more complicated.  However, while such results would certainly be of some theoretical interest, we see no practical drawback to simply working in a particular class of gauges (such as our $P$-smooth class) for which the motion can be sensibly defined and computed.

\section{Inspiral}\label{sec:inspiral}

Our perturbation expansion describes asymptotically small departures from a fixed background metric $g$ and background worldline $\gamma$.  This should allow one to investigate local-in-time effects, such as second-order corrections to quantities already investigated at first order, including gravitational redshift \cite{redshift}, stability of circular orbits \cite{ISCO}, periastron advance \cite{periastron}, loss of energy and angular momentum, and ``snapshot'' waveforms \cite{snapshot}.  However, if the goal is to produce waveforms reflecting an entire inspiral, it is clear that our expansion off of a fixed background geodesic will eventually produce inaccurate results.  In order to produce the waveform templates needed for gravitational-wave data analysis, therefore, it will be necessary to go beyond a perturbation expansion off of a fixed background geodesic.

In principle, it seems clear that one should simply ``patch together'' a sequence of perturbation expansions off of a sequence of background geodesics.  However, the details of implementing such a procedure appear to be quite problematic.  For example, while it seems clear that the new background geodesic should be chosen tangent to the old perturbed motion and that initial data for the new perturbed motion should be trivial, it is far from obvious how to choose the initial data for the new metric perturbation, which satisfies a different field equation (with a different effective source).  The whole procedure is further complicated by the choice of gauge: both the metric perturbation and the position perturbation are gauge-dependent, and one would require a way of ensuring that the new choices are in the ``same'' gauge as the old.  It is easy to see how carelessness in this matter can lead to unphysical results: Since the choice of the next background geodesic depends on the choice of gauge, a naive proposal wherein one simply chooses ``no incoming radiation'' with some gauge choice at each step would produce a final waveform that depends on the gauge choices made.

These difficulties are well-known, and a number of approaches have been developed.  In paper I, we used the Hadamard form (e.g., \cite{poisson-review}) of the Lorenz gauge retarded metric perturbation together with a point particle description to argue that the ``MiSaTaQuWa equation'' \cite{misataquwa}---a modified linearized Einstein equation sourced by a point particle on a non-geodesic trajectory determined by an integrodifferential equation---should provide an accurate long-term description.  Unfortunately this argument has no natural generalization as it stands, since we have given no Hadamard or point particle description at second order.   A derivation of Pound \cite{pound} directly obtains the MiSaTaQuWa equation by expanding in the acceleration of an unspecified worldline, and is a promising route toward obtaining a second-order generalization.  However, both of these approaches depend on the Lorenz gauge in an essential way (through its ``relaxation''), and it has not been investigated whether analogous prescriptions based on relaxing alternative gauge conditions would produce the same physical waveform.  Nevertheless, it seems likely that MiSaTaQuWa equation provides a reliable---if computationally challenging---prescription for first-order-accurate long-term evolution.

An alternative, ``adiabatic'' approach to long-term evolution has been pursued by Mino \cite{mino} and Hinderer and Flanagan \cite{hinderer-flanagan}.  Here, one considers bound orbits of a Kerr black hole and assumes adiabaticity in the sense that the radiation reaction timescale is much longer than the orbital timescale.  This assumption allows one to use self-force results (such as would be provided at second-order by applying the prescription of this paper) to determine an adiabatic evolution of the orbital parameters of the background geodesic.  As in the non-adiabatic approaches, above, the gauge dependence of the prescription has not yet been carefully analyzed.  However, it has been suggested (in both the Mino and the Hinderer-Flanagan approaches) that simple conditions reflecting ``no secular growth over short timescales'' should lead to a gauge-invariant waveform.  If the relevant condition on the gauge can be precisely identified, it should be straightforward to choose such a gauge within our formalism, since we allow a wide class of smoothly-related gauges.  In particular, the Lorenz condition applied to the regular field is a locally defined gauge condition and therefore should lead to perturbations that do not exhibit secular growth.\footnote{By contrast, the approach of \cite{rosenthal} requires one to work in a mass-centered gauge at first order, in which case the metric perturbation should exhibit secular growth.}  Thus the combination of our results with the work of \cite{mino,hinderer-flanagan} appears to be a promising approach to producing second-order-accurate waveform templates for gravitational-wave astronomy of extreme mass-ratio inspirals.

\section{Summary of Prescription}\label{sec:summary}

We conclude by summarizing the prescription for computing the first and second-order motion and metric.  First, choose a vacuum background spacetime $g$, such as Schwarzschild or Kerr.  Next choose a timelike geodesic, $\gamma$, of that spacetime (representing the lowest-order motion of the body), and choose and a point $\gamma_0$ at which the perturbed motion is taken to be coincident.  Determine a coordinate transformation between a global coodinate system for $g$ and a local RWZ coordinate system about the geodesic, equation \eqref{gform}, which in particular determines STF curvature tensors $\{\mathcal{E}_{ij},\mathcal{B}_{ij},\mathcal{E}_{ijk},\mathcal{B}_{ijk}\}$.  Now compute $h^S$ to $O(r^2)$ in terms of these STF tensors from equations \eqref{hS00}-\eqref{hSij} and choose an arbitrary extension, $\hat{h}^S$.    Then compute the effective source, equation \eqref{S1}, and solve equation \eqref{Gs1} for $\hat{h}^R$, imposing a convenient gauge condition on $\hat{h}^R$ such that $\hat{h}^R$ is $C^2$.  The first-order metric perturbation $h$ is then given in a $P$-smooth gauge by $h=\hat{h}^R+\hat{h}^S$.  If one is stopping at first order, one may now determine the first-order motion $Z^{(1)}$ by integrating equation \eqref{Z1hR} with trivial initial data at $\gamma_0$.\footnote{If one is stopping at first order, one only requires that $h^R$ be $C^1$ instead of $C^2$.  Correspondingly, one may choose to compute $h^S$ only to $O(r)$ when constructing the effective source.} 

If one is proceeding to second order, one should instead integrate equations \eqref{transAcoord1}-\eqref{transBcoord2} for $A_\mu$ and $B_{\mu \nu}$. (The integration for $A^\mu$ is redundant with an integration for $Z^{(1)\mu}=A^\mu$ via equation \eqref{Z1hR}.) The initial data for $A_\mu=Z^{(1)}_{\ \ \ \mu}$ and $B_{i0}=\dot{A}_i=\dot{Z}^{(1)}_{\ \ \ i}$ should be trivial (consistent with the interpretation of the particle being initially coincident with the background worldline), while the initial data for $B_{[ij]}$ is arbitrary (trivial being one allowed choice).  Next compute $j^S$ to $O(r)$ from equations \eqref{jS00}-\eqref{jSij} and choose an arbitrary extension, $\hat{j}^S$.  Then compute the second-order effective source, equation \eqref{S2}, and solve equation \eqref{G2s} for $\hat{j}^{R}$, imposing a convenient gauge condition on $\hat{j}^{R}$ (such as the Lorenz condition) such that $\hat{j}^{R}$ is $C^1$.  Finally, the second-order motion $Z^{(2)}$ is given by integrating equation \eqref{Z2hR} with trivial initial data at $\gamma_0$, and the second order metric perturbation $j$ is given by $j=\hat{j}^S+\hat{j}^{R}$.  The first-order motion was previously calculated as $Z^{(1)}_{\ \ \ \mu}=A_\mu$, and the first-order metric perturbation was previously calculated as $h=\hat{h}^R+\hat{h}^S$.  Second-order observables may be constructed from the combination $\{h,Z^{(1)},j,Z^{(2)}\}$.

\begin{acknowledgements}
Support for this research was provided by NSF grant PHY08-54807 to the University of Chicago and by NASA through the Einstein Fellowship Program, grant PF1-120082.  Some of the key ideas for this work arose in conversation with Robert Wald.  The author also acknowledges Abraham Harte for helpful comments.
\end{acknowledgements}

\appendix

\section{The Perturbed Geodesic Equation}\label{sec:pertgeod}

In this appendix we consider the perturbation expansion of a metric and worldline and discuss the gauge freedom through second order.  We then suppose that the worldline is geodesic and derive the perturbative description through second order, expressing the results in the RWZ coordinate system (equation \eqref{gform}) used throughout the paper.

Fix a coordinate system $x^\mu$ on a manifold $M$.  Consider a smooth one-parameter-family of metrics $g_{\mu \nu}(\lambda)$ along with a smooth one-parameter-family of timelike curves $Z^\mu(\lambda;\tau)$.  Taylor expanding the metric components $g_{\mu \nu}(\lambda;x)$ and worldline coordinate position $Z^\mu(\lambda;\tau)$ gives
\begin{align}
g_{\mu \nu}(\lambda; x) & = g^{(0)}_{\mu \nu}(x) + \lambda g^{(1)}_{\mu \nu}(x) + \lambda^2 g^{(2)}_{\mu \nu}(x) + O(\lambda^3) \label{gexp} \\
Z^\mu(\lambda;\tau) & = Z^{(0)\mu}(\tau) + \lambda Z^{(1)\mu}(\tau) + \lambda^2 Z^{(2)\mu}(\tau) + O(\lambda^3), \label{Zexp}
\end{align}
where we have defined metric perturbations $g^{(n)}_{\mu \nu} \equiv (1/n!) \partial_\lambda g_{\mu \nu} |_{\lambda=0}$ and coordinate position peturbations $Z^{(n)\mu} \equiv (1/n!) \partial_\lambda Z^\mu|_{\lambda=0}$ in the usual way.  These quantities depend on the choice of coordinates $x^\mu$.  Under a change $x'^\mu(\lambda;x^\nu)$, the coordinate position $Z^\mu(\lambda;\tau)$ transforms by $Z'^\mu(\lambda;\tau)=x'^\mu(\lambda;z^\mu(\lambda;\tau))$, while the metric components transform via the tensor transformation law.  Restricting to coordinate transformations that reduce to the identity at $\lambda=0$ gives the ``gauge freedom'' within perturbation theory.  Following \cite{pert} we write the coordinate transformation as
\begin{equation}\label{gauge-coord}
 x'^\mu = x^\mu + \lambda \xi^\mu + \frac{1}{2} \lambda^2 \left( \Xi^\mu + \xi^\nu \partial_\nu \xi^\mu \right) + O(\lambda^3),
\end{equation} 
so that the smooth vector fields $\xi$ and $\Xi$ are the first and second-order generators of the diffeomorphism corresponding to the coordinate transformation.  
The transformation laws for $Z^\mu(\lambda)$ and $g_{\mu \nu}(\lambda)$ now give
\begin{align}
Z'^{(1)\mu}(\tau) & = Z^{(1)\mu}(\tau) + \xi^\mu|_{Z^{(0)}(\tau)} \label{Z1t} \\
Z'^{(2)\mu}(\tau) & = Z^{(2)\mu}(\tau) + \Xi^\mu|_{Z^{(0)}(\tau)} \nonumber \\ & + \left. \left[ \left( Z^{(1)\nu}(\tau) + \xi^\nu \right) \partial_\nu \xi^{\nu} \right] \right|_{Z^{(0)}(\tau)} \label{Z2t}
\end{align}
and
\begin{align}
g'^{(1)}_{\mu \nu} & = g^{(1)}_{\mu \nu} - \mathcal{L}_\xi g^{(0)}_{\mu \nu} \label{g1t} \\
g'^{(2)}_{\mu \nu} & = g^{(2)}_{\mu \nu} - \frac{1}{2} \mathcal{L}_\Xi g^{(0)}_{\mu \nu} + \frac{1}{2} \mathcal{L}^2_\xi g^{(0)}_{\mu \nu} - \mathcal{L}_{\xi} g^{(1)}_{\mu \nu}, \label{g2t}
\end{align}
where primed perturbations are defined via Taylor expansion (in $\lambda$) of components in the primed coordinate system.  Under a change of coordinates for the background spacetime (i.e., a $\lambda$-independent change of coordinates for $M$), the metric perturbations transform as tensors on $M$, while the coordinate position perturbations transform as vectors on $Z^{(0)}$.  Thus if we work exclusively within perturbation theory, we may remove $\lambda$ from the description and view the perturbations as tensor fields on the background spacetime that obey additional gauge transformation laws.

Now suppose that each curve $Z^\mu(\lambda;\tau)$ satisfies the affinely-parameterized geodesic equation in $g_{\mu \nu}(\lambda;x)$,
\begin{equation}\label{geod}
\ddot{Z}^\mu + \left. \Gamma^\mu_{\alpha \beta} \right|_{Z(\tau)}\dot{Z}^\alpha \dot{Z}^\beta = 0,
\end{equation}
where an overdot denotes an ordinary derivative with respect to $\tau$ (at fixed $\lambda$).  In perturbing equation \eqref{geod} it is convenient to  choose RWZ coordinates (equation \eqref{gform}) for the background metric, where $Z^{(0)i}=0$ and $Z^{(0)0}=t=\tau$.  Plugging in the expansions \eqref{gexp} and \eqref{Zexp} and collecting powers of $\lambda$ yields
\begin{align}
\ddot{Z}^{(1)}_{\ \ \ 0} & = - \frac{1}{2} \partial_0 g^{(1)}_{00} \nonumber \\
\ddot{Z}^{(1)}_{\ \ \ i} & = - \partial_0 g^{(1)}_{0i} + \frac{1}{2} \partial_i g^{(1)}_{00} - \mathcal{E}_{ij} Z^{(1)j} \label{Z1g}
\end{align}
and
\begin{align}
\ddot{Z}^{(2)}_{\ \ \ 0} & = - \frac{1}{2} \partial_0 g^{(2)}_{00} - g^{(1)}_{0\nu} \ddot{Z}^{(1)\nu} \nonumber \\ & + \dot{Z}^{(1)\gamma} \partial_{\gamma} g^{(1)}_{00} + \frac{1}{2} Z^{(1)\gamma} \partial_\gamma \partial_0 g^{(1)}_{00} \nonumber \\ & - 2 \mathcal{E}_{ij}\dot{Z}^{(1)i}Z^{(1)j} - \frac{1}{2} \dot{\mathcal{E}}_{ij} Z^{(1)i} Z^{(1)j}, \nonumber \\
\ddot{Z}^{(2)}_{\ \ \ i} & = - \partial_0 g^{(2)}_{0i} + \frac{1}{2} \partial_i g^{(2)}_{00} - \mathcal{E}_{ij} Z^{(2)j} \nonumber \\ & - g^{(1)}_{i \nu} \ddot{Z}^{(1)\nu} + Z^{(1)\gamma} \partial_{\gamma} (-\partial_0 g^{(1)}_{0i} + \frac{1}{2} \partial_i g^{(1)}_{00}) \nonumber \\ & + 2 \dot{Z}^{(1)0} \ddot{Z}^{(1)i} - \dot{Z}^{(1)j} \left( \partial_0 g^{(1)}_{ij} + \partial_j g^{(1)}_{i0} - \partial_i g^{(1)}_{j0}\right) \nonumber \\ & -2\dot{Z}^{(1)j}Z^{(1)k} \epsilon_{ijl}\mathcal{B}_k^{\ l} + \frac{2}{3} Z^{(1)k} Z^{(1)l} \epsilon_{ipk} \dot{\mathcal{B}}_l^{\ p} \nonumber \\ & - \frac{1}{2} \mathcal{E}_{ijk} Z^{(1)j} Z^{(1)k} - \dot{\mathcal{E}}_{ij} Z^{(1)0}Z^{(1)j}, \label{Z2g}
\end{align}
where all quantities are evaluated on the background worldline $\gamma$ and the background metric (equal to $\eta_{\mu\nu}$ on $\gamma$) is used to raise and lower indices.  

We have left equations \eqref{Z1g} and \eqref{Z2g} in coordinate form, which is sufficient for practical purposes, since applying the prescription of this paper will require constructing RWZ coordinates in any case.  However, it is straightforward in principle to convert these expressions into covariant language using the formulae for the STF curvature tensors, equations \eqref{Eij}-\eqref{Bijk}, as well as the fact that the background four-velocity is given by $\dot{Z}^{(0)\alpha}=(1,0,0,0)$.  For use in comparing to previous work, we give the covariant version of \eqref{Z1g},
\begin{equation}
\ddot{Z}^{(1)}_{\ \ \ \mu} = \left( -\nabla_\alpha g^{(1)}_{\beta \mu} + \frac{1}{2} \nabla_\mu g^{(1)}_{\alpha \beta} - R^{(0) \ \gamma}_{\alpha \mu \beta} Z^{(1)}_\gamma \right) \dot{Z}^{(0)\alpha} \dot{Z}^{(0)\beta}. \label{Z1cov}
\end{equation}
Equation \eqref{Z1cov} differs from other equations sometimes referred to as the ``perturbed geodesic equation'' in two ways.  The first difference is that we have no projection orthogonal to the background worldline.  This corresponds to our choice of an affine parameter in the perturbed spacetime, equation \eqref{geod}, as opposed to a parameter such that the perturbed tangent vector $\dot{Z}^{(1)\mu}$ is normalized in the background metric.\footnote{Note that in paper I we effectively used the latter parameterization by demanding that our deviation vector be orthogonal to the background worldline.}  The second difference is that a ``geodesic deviation'' term (involving the Riemann tensor of the background) appears in our equation.  If one assumes a $\lambda$-independent metric family, the definitions and calculations of this appendix reproduce standard derivations of the geodesic deviation equation (and provide a second-order generalization).  A version of the perturbed geodesic equation without the geodesic deviation term would have to refer to a definition of the motion perturbation and/or metric peturbation that differs from our straightforward Taylor expansion.

\section{Expressions for the Singular fields}\label{sec:appsing}

Here we display expressions for the first and second order singular fields in the RWZ coordinates for the background metric (equation \eqref{gform}).  For convenience in displaying the results, we have let $M \rightarrow 1$ (corresponding to a choice of units adapted to the small body), so that explicit factors of $M$ do not appear.  Factors of $M$ may be restored on dimensional grounds, and explicit instructions are given below.

The first order singular field $h^S$ is given by equation \eqref{hS} with equations \eqref{M1}, \eqref{a2100}-\eqref{a21ij} and \eqref{aS00}-\eqref{aSij}. Collecting those equations together yields
\begin{align}
h^S_{00} & = \frac{2}{r} + 2 r \mathcal{E}_{ij} n^i n^j + \frac{2}{3} r^2 \mathcal{E}_{ijk} n^i n^j n^k + O(r^3) \label{hS00} \\
h^S_{i0} & = \frac{2}{3} r \epsilon_{ikl} n^j n^k \mathcal{B}_j^{\ l} + \frac{2}{9} r^2 \Big( 2 \epsilon_{ij}^{\ \ m} n^j n^k n^l \mathcal{B}_{klm} \nonumber \\ & \qquad \qquad \qquad + n^j \dot{\mathcal{E}}_{ij} - n_i n^j n^k \dot{\mathcal{E}}_{jk} \Big) + O(r^3) \label{hSi0} \\
h^S_{ij} & = \frac{2}{r} \delta_{ij} - 2r\left(2\mathcal{E}_{ij}+ \delta_{ij} \mathcal{E}_{kl}n^k n^l \right) \nonumber \\ & + \frac{1}{3} r^2 \Big( -4 \epsilon_{kl(i} \dot{\mathcal{B}}_{j)}^{\ l}n^k + 2n_{(i} \epsilon_{j)lm} n^k n^l \dot{\mathcal{B}}_k^{\ m} \nonumber \\ & \qquad - 6 n^k \mathcal{E}_{ijk} -2 \delta_{ij} \mathcal{E}_{klm} n^k n^l n^m \Big) + O(r^3). \label{hSij}
\end{align}
In units where $M \neq 1$, an explicit factor of $M$ would multiply the entire right-hand-sides of the expressions given above.  The second order singular field is given by $j^S = j^P - \mathcal{L}_\xi h^P$ (equation \eqref{jS}), where $j^P$ is given by equations \eqref{jP}, \eqref{M2}, \eqref{a2200}-\eqref{a32ij} and \eqref{deltaExicoord}-\eqref{deltaBxicoord}, $\xi$ is given by equations \eqref{xi0}-\eqref{xii}, and $h^P$ is given by equations \eqref{hP}, \eqref{M1}, \eqref{a2100}-\eqref{aHij} and \eqref{deltaExicoord}-\eqref{deltaBxicoord}.  Computing $j^S = j^P - \mathcal{L}_\xi h^P$ gives
\begin{widetext}
\begin{align}
j^S_{00} & = \frac{-2}{r^2} \Big[ 1 - A^i n_i \Big] - \frac{1}{r} \Big[ 2 h^R_{00} + h^R_{ij} n^i n^j\Big] + \frac{1}{2} \Big[ - 4 \partial^i h^R_{00} n_i - \partial_k h^R_{ij} n^i n^j n^k + 2 A^i n^j \mathcal{E}_{ij} + n^i n^j \mathcal{E}_{ij} + 4 A^i n_i n^j n^k \mathcal{E}_{jk} \Big] \nonumber \\ & + \frac{1}{12} r \Big[ -4 \partial_0 \partial^j h^R_{0i} n_i n_j - 10 \partial_i\partial_j h^R_{00} n^i n^j + 2 \partial_0 \partial_0 h^R_{ij} n^i n^j - 2 \partial_k\partial_l h^R_{ij} n^i n^j n^k n^l - 40 \dot{A}^i \epsilon_{ikl} n^j n^k \mathcal{B}_{j}{}^{l} + 4 A^i \epsilon_{ikl} n^j n^k \dot{\mathcal{B}}_{j}{}^{l} \nonumber \\ & - 4 h^R_{00} n^i n^j \mathcal{E}_{ij} - 16 h^R_{ij} n^i n^j n^k n^l \mathcal{E}_{kl}  +  n^i n^j n^k \mathcal{E}_{ijk} + 8 A^i n_i n^j n^k n^l \mathcal{E}_{jkl}\Big] + O(r^2) \label{jS00} \\
%start jSi0
j^S_{i0} & = \frac{-2}{r} \Big[ h^R_{0i}+2 \dot{A}_i \Big] + \frac{1}{6} \Big[ -12 \partial_i h^R_{0j} n^j + 12 \partial_0 h^R_{ij} n^j + 4 A^j \epsilon_{jkl} n^k \mathcal{B}_{i}{}^{l} + 8 A^j \epsilon_{ikl} n^k \mathcal{B}_{j}{}^{l} -  \epsilon_{ikl} n^j n^k \mathcal{B}_{j}{}^{l} - 8 A^j \epsilon_{ijl} n^k \mathcal{B}_{k}{}^{l} \nonumber \\ & + 4 A^j \epsilon_{ilm} n_j n^k n^l \mathcal{B}_{k}{}^{m}\Big] + \frac{1}{9} r \Big[ -9 \partial_i \partial_k h^R_{0j} n^j n^k + 9 \partial_0 \partial_k h^R_{ij} n^j n^k + 6 \epsilon_{ikm} B^{[lm]} n^j n^k \mathcal{B}_{jl} + 3 \epsilon_{i}^{\ km} h^R_{lm} n^j n^k \mathcal{B}_{jl} \nonumber \\ & - 6 \epsilon_{k}^{\ lm} B_{[il]} n^j n^k \mathcal{B}_{jm} - 3 \epsilon_{k}^{\ lm} h^R_{il} n^j n^k \mathcal{B}_{jm} - 3 \epsilon_{i}^{\ kp} \delta^{lm} h^R_{lm} n^j n_k \mathcal{B}_{jp} + 6 \epsilon_{i}^{\ lm} B_{[jl]} n^j n^k \mathcal{B}_{km} + 3 \epsilon_{i}^{lm} h^R_{jl} n^j n^k \mathcal{B}_{km} \nonumber \\ & - 3 \epsilon_{imp} h^R_{jk} n^j n^k n^l n^m \mathcal{B}_{l}{}^{p} + 8 A^j \epsilon_{jk}{}^{m} n^k n^l \mathcal{B}_{ilm} + 8 A^j \epsilon_{ik}{}^{m} n^k n^l \mathcal{B}_{jlm} - 8 A^j \epsilon_{ij}{}^{m} n^k n^l \mathcal{B}_{klm} - 2 \epsilon_{ij}{}^{m} n^j n^k n^l \mathcal{B}_{klm} \nonumber \\ & + 4 A^j \epsilon_{ik}{}^{p} n_j n^k n^l n^m \mathcal{B}_{lmp} + 36 \dot{A}^j \mathcal{E}_{ij} + 36 \dot{A}^j n_j n^k \mathcal{E}_{ik} + 18 h^R_{0j} n^j n^k \mathcal{E}_{ik} - 18 \dot{A}^j n_i n^k \mathcal{E}_{jk} - 18 h^R_{0i} n^j n^k \mathcal{E}_{jk} + 10 A^j \dot{\mathcal{E}}_{ij} \nonumber \\ & - 76 n^j \dot{\mathcal{E}}_{ij} + 10 A^j n_j n^k \dot{\mathcal{E}}_{ik} - 2 A^j n_i n^k \dot{\mathcal{E}}_{jk} + 2 A_i n^j n^k \dot{\mathcal{E}}_{jk} - 17 n_i n^j n^k \dot{\mathcal{E}}_{jk} - 2 A^j n_i n_j n^k n^l \dot{\mathcal{E}}_{kl} \Big] + O(r^2) \label{jSi0} \\
%start jSij
j^S_{ij} & = \frac{1}{2r^2} \Big[3 \delta_{ij} + 4 A^k \delta_{ij} n_k\Big] + \frac{1}{r} \Big[2 h^R_{ij} - \delta_{ij} h^R_{kl} n^k n^l\Big] + 2 \partial_k h^R_{ij} n^k - \frac{1}{2} \delta_{ij} \partial_m h^R_{kl} n^k n^l n^m - 8 \mathcal{E}_{ij} + 4 A^k n_k \mathcal{E}_{ij} + A^k \delta_{ij} n^l \mathcal{E}_{kl} \nonumber \\ & + \delta_{ij} n^k n^l \mathcal{E}_{kl} - 2 A^k \delta_{ij} n_k n^l n^m \mathcal{E}_{lm} + \frac{1}{24} r \Big[ 8 \partial_i\partial_j h^R_{00} - 8 \partial_0 \partial_j h^R_{0i} - 8 \partial_0 \partial_i h^R_{0j} + 8 \partial_0 \partial_0 h^R_{ij} + 8 \partial_0 \partial^k h^R_{0j} n_i n_k \nonumber \\ & + 8 \partial_0 \partial_j h^R_{0k} n_i n^k + 8 \partial_0 \partial^k h^R_{0i} n_j n_k + 8 \partial_0 \partial_i h^R_{0k} n_j n^k - 8 \partial_j\partial_k h^R_{00} n_i n^k - 8 \partial_0 \partial_0 h^R_{jk} n_i n^k - 8 \partial_i\partial_k h^R_{00} n_j n^k \nonumber \\ &  - 8 \partial_0 \partial_0 h^R_{ik} n_j n^k - 72 \delta_{ij} \partial_0 \partial^l h^R_{0k} n_k n_l + 48 \partial_0 \partial^l h^R_{0k} n_i n_j n_k n_l + 36 \delta_{ij} \partial_k\partial_l h^R_{00} n^k n^l + 36 \delta_{ij} \partial_0 \partial_0 h^R_{kl} n^k n^l \nonumber \\ & + 16 \partial_k\partial_l h^R_{ij} n^k n^l + 8 \partial_j\partial_l h^R_{ik} n^k n^l + 8 \partial_i\partial_l h^R_{jk} n^k n^l - 8 \partial_i\partial_j h^R_{kl} n^k n^l - 24 \partial_k\partial_l h^R_{00} n_i n_j n^k n^l - 24 \partial_0 \partial_0 h^R_{kl} n_i n_j n^k n^l \nonumber \\ & - 4 \delta_{ij} \partial_m\partial_p h^R_{kl} n^k n^l n^m n^p - 16 \dot{A}^k \epsilon_{jkl} \mathcal{B}_{i}{}^{l} - 16 \dot{A}^k \epsilon_{klm} n_j n^l \mathcal{B}_{i}{}^{m} + 16 \dot{A}^k \epsilon_{jlm} n_k n^l \mathcal{B}_{i}{}^{m} + 16 \epsilon_{jlm} h^R_{0k} n^k n^l \mathcal{B}_{i}{}^{m} \nonumber \\ & - 16 \dot{A}^k \epsilon_{ikl} \mathcal{B}_{j}{}^{l} - 16 \dot{A}^k \epsilon_{klm} n_i n^l \mathcal{B}_{j}{}^{m} + 16 \dot{A}^k \epsilon_{ilm} n_k n^l \mathcal{B}_{j}{}^{m} + 16 \epsilon_{ilm} h^R_{0k} n^k n^l \mathcal{B}_{j}{}^{m} + 16 \dot{A}^k \epsilon_{jkm} n_i n^l \mathcal{B}_{l}{}^{m} \nonumber \\ & + 16 \dot{A}^k \epsilon_{ikm} n_j n^l \mathcal{B}_{l}{}^{m} + 144 \dot{A}^k \epsilon_{kmp} \delta_{ij} n^l n^m \mathcal{B}_{l}{}^{p} - \dot{A}^k \epsilon_{kmp} n_i n_j n^l n^m \mathcal{B}_{l}{}^{p} + 12 A^k \epsilon_{jkl} \dot{\mathcal{B}}_{i}{}^{l} - 32 \epsilon_{jkl} n^k \dot{\mathcal{B}}_{i}{}^{l} \nonumber \\ & - 4 A^k \epsilon_{klm} n_j n^l \dot{\mathcal{B}}_{i}{}^{m} + 20 A^k \epsilon_{jlm} n_k n^l \dot{\mathcal{B}}_{i}{}^{m} + 12 A^k \epsilon_{ikl} \dot{\mathcal{B}}_{j}{}^{l} - 32 \epsilon_{ikl} n^k \dot{\mathcal{B}}_{j}{}^{l} - 4 A^k \epsilon_{klm} n_i n^l \dot{\mathcal{B}}_{j}{}^{m} + 20 A^k \epsilon_{ilm} n_k n^l \dot{\mathcal{B}}_{j}{}^{m} \nonumber \\ & - 4 A_j \epsilon_{ilm} n^k n^l \dot{\mathcal{B}}_{k}{}^{m} - 4 A_i \epsilon_{jlm} n^k n^l \dot{\mathcal{B}}_{k}{}^{m} + 35 \epsilon_{jlm} n_i n^k n^l \dot{\mathcal{B}}_{k}{}^{m} + 35 \epsilon_{ilm} n_j n^k n^l \dot{\mathcal{B}}_{k}{}^{m} + 4 A^k \epsilon_{jkm} n_i n^l \dot{\mathcal{B}}_{l}{}^{m} \nonumber \\ & + 4 A^k \epsilon_{ikm} n_j n^l \dot{\mathcal{B}}_{l}{}^{m} + 48 A^k \epsilon_{kmp} \delta_{ij} n^l n^m \dot{\mathcal{B}}_{l}{}^{p} - 32 A^k \epsilon_{kmp} n_i n_j n^l n^m \dot{\mathcal{B}}_{l}{}^{p} + 8 A^k \epsilon_{jmp} n_i n_k n^l n^m \dot{\mathcal{B}}_{l}{}^{p} \nonumber \\ & + 8 A^k \epsilon_{imp} n_j n_k n^l n^m \dot{\mathcal{B}}_{l}{}^{p} - 16 h^R_{00} \mathcal{E}_{ij} - 64 h^R_{kl} n^k n^l \mathcal{E}_{ij} - 96 B_{[jk]} \mathcal{E}_{i}^{\ k} - 48 h^R_{jk} \mathcal{E}_{i}^{\ k} + 16 h^R_{00} n_j n^k \mathcal{E}_{ik} + 16 h^R_{jk} n^k n^l \mathcal{E}_{il} \nonumber \\ & - 96 B_{[ik]} \mathcal{E}_{j}^{\ k} - 48 h^R_{ik} \mathcal{E}_{j}^{\ k} + 16 h^R_{00} n_i n^k \mathcal{E}_{jk} + 16 h^R_{ik} n^k n^l \mathcal{E}_{jl} - 72 \delta_{ij} h^R_{00} n^k n^l \mathcal{E}_{kl} - 16 h^R_{ij} n^k n^l \mathcal{E}_{kl} + 48 h^R_{00} n_i n_j n^k n^l \mathcal{E}_{kl} \nonumber \\ & + 96 \delta_{ij} B_{[km]} n^k n^l \mathcal{E}_{l}^{m} + 48 \delta_{ij} h^R_{km} n^k n^l \mathcal{E}_{l}^{\ m} - 96 B_{[km]} n_i n_j n^k n^l \mathcal{E}_{l}^{\ m} - 48 h^R_{km} n_i n_j n^k n^l \mathcal{E}_{l}^{\ m} \nonumber \\ & + 16 \delta_{ij} h^R_{kl} n^k n^l n^m n^p \mathcal{E}_{mp} - 96 A_0 \dot{\mathcal{E}}_{ij} + 48 A_0 \delta_{ij} n^k n^l \dot{\mathcal{E}}_{kl} - 48 A_0 n_i n_j n^k n^l \dot{\mathcal{E}}_{kl} + 48 A^k \mathcal{E}_{ijk} - 96 n^k \mathcal{E}_{ijk} + 48 A^k n_k n^l \mathcal{E}_{ijl} \nonumber \\ & - 48 A^k \delta_{ij} n^l n^m \mathcal{E}_{klm} + 48 A^k n_i n_j n^l n^m \mathcal{E}_{klm} + 6 \delta_{ij} n^k n^l n^m \mathcal{E}_{klm} - 16 A^k \delta_{ij} n_k n^l n^m n^p \mathcal{E}_{lmp}\Big] + O(r^2) \label{jSij}
\end{align}
\end{widetext}
In units where $M \neq 1$, an explicit factor of $M$ would multiply the terms involving $h^R$, $A$ and $B$, while an explicit factor of $M^2$ would multiply the remaining terms.  In writing the above results we have used equations \eqref{fixB} and \eqref{transAcoord2} to express $B_{\mu \nu}$ entirely in terms of $h^R_{\mu \nu}$, $\dot{A}^i$ and $B_{[ij]}$.  Thus the second-order singular field depends on the quantities $\{A^\mu, B_{[ij]}, h^R_{\mu \nu}, \mathcal{E}_{ij}, \mathcal{B}_{ij}, \mathcal{E}_{ijk}, \mathcal{B}_{ijk} \}$.  We remind the reader that $A^\mu$ represents the first-order motion, $A^\mu=Z^{(1)\mu}$, so that the second-order singular field may be viewed as depending on the background metric (through $\{\mathcal{E}_{ij}, \mathcal{B}_{ij}, \mathcal{E}_{ijk}, \mathcal{B}_{ijk}\}$), the first-order metric (through the regular field $h^R=h-h^S$), as well as choices of initial data for the first-order motion, $Z^{(1)\mu}$, and relative spatial coordinate alignment, $B_{[ij]}$.

 Equations \eqref{hS00}-\eqref{jSij} for the first and second-order singular fields are the main computational results of this paper.  We have performed and verified the calculations leading to these expressions using the tensor analysis package \textit{xTensor} \cite{xTensor} for the software package \textit{Mathematica} \cite{mathematica}.

\end{document}